\documentclass[Journals,InsideFigs,NoLineNumbers]{ascelike}

\usepackage{graphicx}
\usepackage{subfigure}
\usepackage{amsmath}
\usepackage{amsfonts}
\usepackage{amssymb}
\usepackage{amsbsy}
\usepackage{times}
\usepackage[utf8]{inputenc}
\usepackage[T1]{fontenc}
\usepackage{lmodern}
\usepackage{newtxtext,newtxmath}
\usepackage[noblocks]{authblk}
\usepackage{natbib}
\usepackage[figurename=Fig.,labelfont=bf,labelsep=period]{caption}
\newcommand{\Expect}{{\rm I\kern-.3em E}}
\newcommand{\Var}{\mathrm{Var}}
\providecommand{\abs}[1]{\lvert#1\rvert}

\begin{document}

% You will need to make the title all-caps
\title{Reliability-Based Collapse Assessment of Wind-Excited Steel Structures within Performance-Based Wind Engineering
\let\thefootnote\relax\footnotetext
{S. Arunachalam and S.M.J. Spence. Forthcoming. "Reliability-Based Collapse Assessment of Wind-Excited Steel Structures within Performance-Based Wind Engineering." ASCE Journal of Structural Engineering. 10.1061/(ASCE)ST.1943-541X.0003444. 
}}

\author{Srinivasan Arunachalam, S.M.ASCE}
\affil{Graduate Student, Department of Civil and Environmental Engineering, University of Michigan, Ann Arbor, MI 48109. E-mail: sriarun@umich.edu}
\author{Seymour M.J. Spence, Ph.D., A.M.ASCE}
\affil{Associate Professor, Department of Civil and Environmental Engineering, University of Michigan, Ann Arbor, MI 48109 (corresponding author). E-mail: smjs@umich.edu}

\maketitle

% Please include an abstract:
\begin{abstract}

As inelastic design for wind is embraced by the engineering community, there is an increasing demand for computational tools that enable the investigation of the nonlinear behavior of wind-excited structures and subsequent development of performance criteria. To address this need, a probabilistic collapse assessment framework for steel structures is proposed in this paper. The framework is based on the integration of a high-fidelity fiber-based nonlinear structural modeling environment with a wind-tunnel-informed stochastic wind load model to perform nonlinear time history analysis. General uncertainty is propagated using a stratified sampling scheme enabling the efficient estimation of reliabilities associated with rare events. The adopted models for simulating high-fidelity nonlinear structural behavior were found, in general, to be adequate for capturing phenomena, including progressive yielding, buckling, and low-cycle fatigue, that are essential for wind induced collapse analysis. In particular, the adopted fatigue model was found to be capable of predicting damage and potential fiber/section fracture associated with non-fully reversing stress-strain cycles that are characteristic of wind loading. Through illustration on a 45-story archetype steel building, critical discussions on the types of observed collapse mechanisms, the difference between alongwind and acrosswind nonlinear behavior, reliabilities associated with first yield, and collapse are presented. A probabilistic description of the residual and peak story drifts is also provided through development of fragility functions.

\end{abstract}


\section{Introduction}
\label{sec:intro}  
  
More than a decade of active research in performance-based wind engineering (PBWE) has led to the development of several frameworks \citep{Jain_01,Ciampoli_2011,Petrini_12,barbato2013,Bernardini2015,Chuang_17,Cui_18,ZhiCheng20} to facilitate the performance assessment of structures at various wind intensities. This work has demonstrated the promise of PBWE as an effective means to shift away from conventional design for wind that is predicated on elastic analysis and equivalent static loads. This concerns the design of both new structures, \citep{Chuang_17,Judd2,ZhiCheng20}, as well as rehabilitation or retrofit of existing structures \citep{Muthukumar,Larry_19}. PBWE has been recognized as a national research priority \citep{CTBUH14} with research towards the adoption of performance-based techniques for wind culminating in the release by the American Society of Civil Engineers (ASCE) of the Prestandard for Performance-Based Wind Design (PBWD) \citep{Prestandard19}.

Current design practice for wind is based on the satisfaction of a number of highly prescriptive design requirements at various levels of hazard intensity, e.g., serviceability to life safety. In particular, satisfactory performance at ultimate load levels (i.e. at life safety) is deemed to be achieved through the satisfaction of a series of component-level elastic acceptance criteria at wind demand levels, e.g., wind demands with 700 to 3000 year mean recurrence intervals \citep{ASCE7}, that are well below those for which significant inelasticity and/or collapse is expected to occur. Therefore, when using traditional approaches for wind design (as well as other natural hazards), the actual performance of the system is never explicitly evaluated, it is presumed to be achieved through the satisfaction of a series of component-level prescriptions set at lower load intensities. Performance-based engineering is a fundamental paradigm shift in which performance is explicitly evaluated at system-level for each limit state of interest (from serviceability to life safety/collapse). These methods are gaining significant traction in practice as better structural performance (through knowledge and control of post-elastic structural behavior) and improved economy can be achieved. This also fosters greater innovation in design as the designer has considerable latitude in demonstrating acceptable building performance, especially in multi-hazard environments with comparable wind and seismic demands. This interest from practice for a shift towards a PBWD philosophy is evident from the publication of the ASCE Prestandard on PBWD \citep{Prestandard19}. Importantly, this document recognizes how, to achieve the maximum benefits from a PBWD, ultimate performance should be evaluated explicitly through a system-level collapse analysis with uncertainty treated through reliability. The goal of this study is to outline a framework that enables such an approach.    

Indispensable to such an analysis is the modeling of nonlinear dynamic structural effects. Although the analysis procedures developed for the implementation of performance-based seismic design \citep{FEMA_P_58,peertall2017} can be useful in developing guidelines for nonlinear dynamic analysis in PBWE, there are fundamental differences in the nature of the structural demands and responses that precludes the simple transition of knowledge. In combination with inelasticity, the long duration of wind loading and the significant mean load component for certain wind directions (alongwind type loading) require explicit modeling of phenomena associated with potential failure due to ratcheting, low-cycle fatigue (LCF) and fracture, especially in the vicinity of collapse. Notwithstanding the difficulties associated with nonlinear structural modeling, considerable research efforts in the past have considered various structural modeling approaches, including static and dynamic single-/multi-degree-of-freedom models with both concentrated and distributed plasticity \citep{Beck_14,Tabbuso_16,Chuang_17,Judd2,Feng_17,Feng_18,Larry_19,Chuang_19,Chuang_20,Ouyang_21,Ghaffary_21,Huang_2022,Chuang_2022}. 
The primary limitation of concentrated plasticity models is that the accuracy is predicated on the chosen parameters for defining the hysteretic force-deformation relationships of the hinges. Fiber-based models explicitly account for the growth of plastification along the length, width and depth of a member. The success of fiber-based models is dependent only on the correctness of the uniaxial fiber stress-strain relationship. The ability to explicitly model axial-moment interaction, and to account for shifting neutral axis as the section begins yielding, makes it a preferred modeling choice. Among the aforementioned studies, different types of analyses, namely, incremental dynamic analysis (IDA) \citep{Judd0,Larry_19}, nonlinear pushover analysis \citep{Judd2,Larry_19}, and nonlinear time history analysis (NLTHA) \citep{Judd1,Judd2,Muthukumar,Larry_19,Ouyang_21,Ghaffary_21,Huang_2022} have been carried out. The IDA samples only cater to collapse assessment and great care must be taken in choosing the wind records to use in the analysis to ensure the proper consideration of wind directionality and record-to-record variability (wind load stochasticity). The significant limitation of pushover analyses is the inability to account for dynamic amplification of displacements, damping effects, and damage accumulation mechanisms resulting from ratcheting and LCF. More recently, \citet{Chuang_17,Chuang_19,Chuang_20,Chuang_2022} proposed a probabilistic performance assessment framework using dynamic shakedown which has been recommended in Appendix C (``Method 3'') of the ASCE Prestandard on PBWD. The framework is successful in reducing the computational demands that NLTHA incurs but is hinged on the description of safety against collapse through the occurrence of the state of dynamic shakedown. Since a structure beyond the state of shakedown is not necessarily unsafe, the analysis yields a conservative estimate of the collapse-level reliability. 
Concerning recent approaches based on NLTHA, \cite{Ghaffary_21} conducted a performance-based assessment of a 20-story building through collapse and studied the nonlinear structural response of steel frames under alongwind and acrosswind loads separately. In conducting the collapse simulation, the frames were subjected to steadily increasing wind loads until instability. The study highlighted the potential benefits, in terms of increased economy and safety, of design for controlled inelasticity under extreme winds. More lately, \cite{Huang_2022} investigated the inelastic performance of high-rise buildings to simultaneous action of alongwind and acrosswind loads. The influence of yielding on the development of drift, and the second-order P-Delta effect on both elastic and inelastic responses were also examined. The study emphasized the importance of considering alongwind and acrosswind loads simultaneously, and the potential significance of P-Delta effects in the response of flexible buildings. 

Central to state-of-the-art PBWE is the explicit propagation of uncertainty during the estimation of the performance metrics \citep{Chuang_17,Cui_18,ierimonti2019cost,Micheli_19,ZhiCheng20,Cui_2020,Ouyang_21}. Such a treatment entails both probabilistic wind load modeling, including the stochasticity of the dynamic wind loads, as well as characterization of the model uncertainty (e.g. uncertainties in the material/mechanical properties, geometric imperfections). To date, with the exception of the approaches outlined in \citet{Chuang_17,Chuang_19,Chuang_2022} that are built around the need to explicitly propagate uncertainty at the cost of a more general description of collapse, few methodologies have been proposed in the literature for characterizing the collapse performance of wind excited systems while comprehensively treating uncertainty through reliability. 

In response to these needs, a general framework for performing a full reliability assessment of wind-excited steel systems experiencing inelasticity until collapse is proposed in this study. A high-fidelity nonlinear modeling environment is adopted to adequately simulate the structural dynamic behavior including yielding, buckling, and LCF-induced fracture until potential collapse. The load modeling entails joint probabilistic modeling of the site-specific wind speeds and directions as well as the stochastic modeling of the dynamic wind loads. A full range of structural and load uncertainties are propagated through the system using a recently introduced stratified sampling scheme \citep{esrel21}. The applicability of the proposed framework is illustrated on a steel braced frame. The results are discussed and key insights into the collapse behavior of steel structures subject to extreme winds are gained. Although this study is not concerned with the development of performance assessment frameworks for structures subject to multiple hazards, the first step in the development of such frameworks is the enrichment of existing PBWE methodologies for the treatment of a comprehensive range of performance objectives. This study contributes to this vision through the development of a PBWE framework for collapse assessment under extreme winds.

%==================================
\section{Outline of proposed framework}
\label{sec:Outline_PF}
\subsection{Reliability Estimation}
\label{sec:PS}

Current wind design for the strength requires component-level limit states, evaluated using elastic analysis and equivalent static wind loads, to be met, corresponding to a specified hazard intensity based on the building risk category. In contrast, advanced PBWE for the strength-level design seeks to ensure minimum system-level reliability against collapse over a given lifespan through an explicit probabilistic evaluation of the structural demands and capacities, preferably using nonlinear dynamic analyses, under a comprehensive set of wind load and model uncertainties. 
Such an approach entails numerically solving the following probabilistic integral, similar to the one discussed in \citet{ZhiCheng20}:
\begin{equation}
\label{eq:probintg}
\lambda_\mathcal{R} = \int_{\alpha} \int_{\bar{v}_H} G(\mathcal{R}|\alpha, \bar{v}_H) \abs{dG(\alpha|\bar{v}_H)} \abs{d\lambda(\bar{v}_H)} 
\end{equation}
where $\mathcal{R}$ is a general explicit or implicit limit state whose exceedance indicates the occurrence of a failure state of interest (e.g. component buckling, component fatigue failure, system collapse); $\lambda_\mathcal{R}$ is the annual rate of exceedance of the limit state $\mathcal{R}$; $G$ is the complementary cumulative distribution function (CCDF); $\bar{v}_H$ is the maximum mean-hourly wind speed measured at the building top height $H$; $\alpha$ is the direction of the impinging wind; $\lambda(\bar{v}_H) $ is the non-directional hazard curve representing the annual rate of exceeding $\bar{v}_H$.  
Eq. (\ref{eq:probintg}) is a statement of total probability that explicitly captures the stochastic nature of the wind loads in addition to the aleatory and epistemic uncertainties in model parameters and and gravity loads through $G(\mathcal{R}|\alpha, \bar{v}_H)$. Vital to the application of Eq. (\ref{eq:probintg}) to high-fidelity analysis, is the adoption of efficient stochastic simulation schemes that enables the estimation of probabilities, and therefore reliabilities, associated with rare events, e.g. collapse.   

For small probabilities, $\lambda_\mathcal{R}$ converges to the annual failure probability $P^a_{f_\mathcal{R}}$ \citep{ Kiureghian_PEER}. This convergence can be used to directly estimate the reliability index, $\beta_T$, associated with the limit state $\mathcal{R}$ as:
\begin{equation}
\label{eq:Reliability_eq}
\beta_T = \Phi^{-1}[(1-P^a_{f_\mathcal{R}})^T] 
\end{equation}
where $\Phi$ is the standard normal distribution function while $T$ is the reference period over which $\beta_T$ is estimated, e.g. 50 years.    

% ------------------------------------------------------------
\subsection{Wind Load and Structural Model Requirements}
\label{sec:modelreqts}

To enable the development of a stochastic simulation-based framework for the efficient estimation of collapse performance through solving Eq. (\ref{eq:probintg}), appropriate high-fidelity wind load and nonlinear structural models must first be identified and subsequently integrated together. 

With respect to the wind loads, the adopted model must first possess the capability to describe the occurrence and intensity of site-specific wind events through the joint probability distribution of wind speed and direction. This will enable the direct estimation of the non-directional hazard curve, $\lambda(\bar{v}_H)$, and the associated CCDF of wind direction given site-specific wind speed, $G(\alpha|\bar{v}_H)$. In addition, models are required for the simulation of stochastic wind load histories for any given realization of the intensity pair, $(\bar{v}_H, \alpha)$. Because of the focus of this work on high-rise structures, the wind load time history model must be capable of capturing building-specific aerodynamic features, to the extent captured in wind tunnel tests, as well as modeling the record-to-record variability inherent to wind excitation.

With respect to nonlinear structural modeling, the modeling environment must allow for the computation of highly nonlinear responses generated by the application of dynamic wind load histories. In general, this will require solving the following equation of motion:
\begin{equation}
\label{eq:EOMeq}
\textbf{M}\ddot{\textbf{u}}(t)+ \textbf{f}_D (\dot{\textbf{u}}(t),\textbf{u}(t)) + \textbf{f}_r (\textbf{u}(t)) = \textbf{f}(t;\bar{v}_H,\alpha)
\end{equation}
where $\textbf{u},\dot{\textbf{u}}$ and $\ddot{\textbf{u}}$ represent the vector of displacements, velocities, and accelerations at the discretized degrees of freedom at time $t$; $\textbf{M}$ is the mass matrix; $\textbf{f}_D$ and $\textbf{f}_r$ are the vectors of damping and restoring forces at time $t$ which have nonlinear dependence on $\textbf{u}$ and encapsulate the effects of material and geometric nonlinearity in the analysis; while $\textbf{f}(t;\bar{v}_H,\alpha)$ is a realization of the external aerodynamic loads, estimated from the wind load model for a given wind speed and direction pair, i.e., $(\bar{v}_H, \alpha)$. In general, the modeling of collapse requires the consideration of multiple component level limit states, the progressive exceedance of which, can ultimately lead to collapse. Therefore, the nonlinear modeling environment must be capable of capturing phenomena such as: component-level first yield; system-level first yield; component fracture due to low-cycle fatigue; and global component buckling. Naturally, the required capability to simulate relevant responses with high accuracy dictates the complexity of the adopted structural modeling environment. Concerning the propagation of uncertainty, Eq. (\ref{eq:EOMeq}) can be viewed as a stochastic differential equation embedding a wide range of structural model (collected in the vector $\boldsymbol{\xi}_m$) and load uncertainties characterized by: $\bar{v}_H$ and $\alpha$; gravity load uncertainties, $\boldsymbol{\xi}_g$; wind load model uncertainties, $\boldsymbol{\xi}_w$; and stochastic noise, $\boldsymbol{\phi}$, generating the stochasticity, or record-to-record variability, in the external dynamic loads $\textbf{f}$. %collected in the vector

\begin{figure}[!b] %!htb
    \centering
	\includegraphics[scale=0.65]{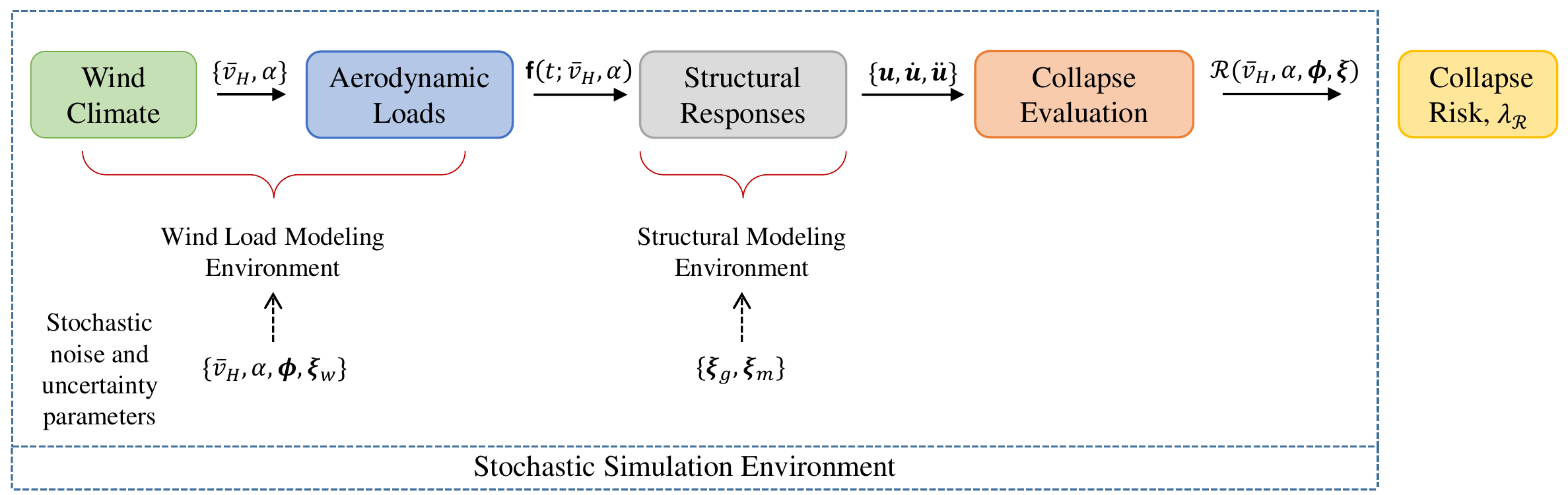}
	\caption{The model chain and uncertainty propagation problem of the collapse assessment framework.}
	\label{fig:propframework}
\end{figure}

In the following section, a candidate wind load model and fiber-based structural modeling environment will be briefly presented. These state-of-the-art models are central ingredients of the proposed framework and have been included with careful consideration of their capabilities to meet the above discussed requirements. Fig. \ref{fig:propframework} schematically represents the proposed framework and highlights the model chain and uncertainty propagation problem (where $\boldsymbol{\xi} = \{\boldsymbol{\xi}_w,\boldsymbol{\xi}_g,\boldsymbol{\xi}_m\}$). 

% ===========================================

% ==========================================

Concerning the overall applicability of the proposed framework, it is important to observe how recent research on integrated wind-induced structural and envelope damage assessment reported that the envelope was seen to be primarily intact for wind events with annual rates as low as $1\times 10^{-6}$ \citep{Ouyang_21}. Inelasticity (including P-Delta effects \citep{Ouyang_21}) in the structural system is not therefore expected to lead to cladding damage that will substantially affect the applied aerodynamic loads. Having said this, with increasing structural damage, particularly near collapse, cladding damage will eventually affect the spatial net pressure/force distribution. The neglect of this damage to the building envelope forms an important assumption of the proposed framework. While the influence of the building envelope damage on the collapse reliability of wind excited system, and its correlation with structural damage near collapse, is outside the scope of this study, it certainly merits future investigation.

% ------------------------------------------------------------

\section{Background models}
\label{Sec:Backgroundmodels}

\subsection{Wind Load Model}
\label{Sec:windmodel}

\subsubsection{Hazard curve and wind directionality}
\label{Sec:hazcurve_dir}

The joint distribution of wind speed and direction can be estimated from site-specific directional wind speed data. It is assumed that the wind events conform to a classic hurricane hazard description where straight (i.e. constant direction) and stationary wind events of 1-hour duration are considered. The relation between the marginal distribution for wind speed and the non-directional hazard curve, $\lambda(\bar{v}_H)$, is expressed in the following form:
\begin{equation}
\label{eq:hazardeq}
\lambda(\bar{v}_H) = \nu (1-F_{\bar{V}_H}(\bar{v}_H))
\end{equation}
where $\nu$ is the annual occurrence rate of storms at the site of interest, obtained from meteorological/simulated data, and $F_{\bar{V}_H}$ is the cumulative distribution function (CDF) of $\bar{v}_H$ given the occurrence of a storm. $F_{\bar{V}_H}$ can be estimated by fitting an appropriate extreme value distribution to site-specific non-directional wind speed data. In general, these wind speeds are obtained from raw data collected at nearby meteorological stations after transformation to obtain a target averaging time (e.g., 1 hour), site-specific terrain roughness, and height (e.g., building height). A probabilistic approach can be adopted for this transformation that systematically accounts for uncertainties in roughness lengths, conversion between averaging times, modeling errors, as well as sampling and observational errors in raw wind speed \citep{Minciarelli2001,Diniz2004,Spence2014}. A detailed discussion on the adopted transformation, along with the associated uncertainties and their distributions, can be found in \citet{Diniz2004}. Throughout the paper, the vector collecting the parameters modeling the aforementioned uncertainties will be indicated as $\textbf{e}$. %Corresponding to each realization of the vector $\textbf{e}$, paired samples $\{\bar{v}_H^j, \alpha ^j\}$, can be obtained, for $j = 1,2,\ldots,N_s$ where $N_s$ is the number of storms in the wind speed data set. 

The dependence between wind speed and direction, as observed in the available wind speed data, is captured through the term $G(\alpha|\bar{v}_H)$ which is given by: 
\begin{equation}
\label{eq:condpdeq}
G(\alpha|\bar{v}_H) = \int_{\alpha}^{2\pi} \frac{f_{\alpha,\bar{V}_H}(\theta,\bar{v}_H)}{f_{\bar{V}_H}(\bar{v}_H)} d\theta
\end{equation}
where $f_{\alpha,\bar{V}_H} $ is joint probability density function (PDF) of $\bar{v}_H$ and $\alpha$ given the occurrence of a storm while $f_{\bar{V}_H} $ is the PDF of $\bar{v}_H$ given a storm. The integrand in Eq. (5) is the conditional PDF of $\alpha$ given $\bar{v}_H$, the integration of which in the range $\theta \in (\alpha, 2\pi)$, yields the conditional CCDF, $G(\alpha|\bar{v}_H)$, of $\alpha$ given $\bar{v}_H$. As outlined in \citet{ZhiCheng20}, a numerical approach can adopted to estimate $f_{\alpha,\bar{v}_H}$ and subsequently $G(\alpha|\bar{v}_H)$. Bivariate copula-based modeling is central to this approach, allowing the marginal distribution models to be uncoupled from the dependence modeling through a bivariate copula density function, $c_{\alpha,\bar{V}_H}$. This allows the joint PDF between $\bar{v}_H$ and $\alpha$ to be written as the product of their marginal PDFs and the copula density, and therefore as:
\begin{equation}
\label{eq:copuladen}
f_{\alpha,\bar{V}_H}(\theta,\bar{v}_H) = c_{\alpha,\bar{V}_H}(F_{\alpha}(\theta),F_{\bar{V}_H}(\bar{v}_H))f_{\alpha}(\theta)f_{\bar{V}_H}(\bar{v}_H)
\end{equation}
This relation allows Eq. (\ref{eq:condpdeq}) to be rewritten in the form:
\begin{equation}
\label{eq:condpdeq2}
G(\alpha|\bar{v}_H) = \int_{\alpha}^{2\pi} c_{\alpha,\bar{v}_H}(F_{\alpha}(\theta),F_{\bar{V}_H}(\bar{v}_H)) f_{\alpha}(\theta)d\theta
\end{equation}
where $F_{\alpha}$ and $f_{\alpha}$ are the CDF and PDF of $\alpha$ given the occurrence of a storm. In view of the bounded nature of $\alpha$, a circular kernel density should be adopted in estimating $f_{\alpha}$ from the directional wind speed data. A specific class of nonparametric density estimators, denominated kernel estimator, is used in this work to model the copula density function in Eq. (\ref{eq:condpdeq2}) \citep{kdecopula2018}. 

% -----------------

\subsubsection{Load stochasticity}
\label{Sec:PODmodel}

In general, to characterize the vector of aerodynamic loads, $\textbf{f}(t)$, acting on a high-rise or tall building, specific boundary layer wind tunnel tests are carried out for a number of wind directions. This ensures the capture, to the extent possible in boundary layer wind tunnel, in the dynamic wind loads of phenomena such as vortex shedding, detached flow and interference effects from surrounding buildings for a full set of wind directions. Through classic scaling procedures (e.g., Strouhal number matching), the model scale loads can be transferred to full scale and calibrated to any wind direction, $\alpha$, and wind speed, $\bar{v}_H$, of interest. While this procedure is very well understood and widely used in practice, it does not capture the record-to-record variability, stochasticity, of the dynamic wind loads $\textbf{f}(t)$. Because of the path dependent nature of nonlinear dynamic analysis, this uncertainty requires consideration.           

To this end, the data-driven proper-orthogonal decomposition (POD)-based spectral representation model outlined in \citet{ArthriyaPOD} is adopted in this work. For each wind direction of interest, this model is based on estimating the cross power spectral density (XPSD) matrix, $\textbf{S}_{\textbf{f}}$, directly from the wind tunnel realization of $\textbf{f}(t)$. From the spectral eigenvalues and eigenvectors of $\textbf{S}_{\textbf{f}}$, the fluctuating component of $\textbf{f}(t)$ can be decomposed into      
a sum of vector valued subprocesses that are noncoherent, enabling independent simulation. In particular, decomposing $\textbf{f}(t)$ into $N_l$ zero mean subprocesses, the $i$th component of $\textbf{f}(t)$ can be written as:    
\begin{equation}
\label{eq:PODeq1}
f_i(t;\bar{v}_H,\alpha) = \bar{f}_i(\bar{v}_H,\alpha) + 2\sum_{l=1}^{N_l} \sum_{j=0}^{N_{\omega}-1} |\Psi_{il}(\omega_j;\alpha)| \sqrt{\Lambda_l(\omega_j;\bar{v}_H,\alpha)\Delta\omega}
\cos(\omega_jt + \theta_{il}(\omega_j) + \phi_{jl})
\end{equation}
where $\bar{f}_i$ is the $i$th component of the mean wind load (estimated directly from the wind tunnel data); $\Delta\omega$ is the frequency increment; $\omega_j = j\Delta\omega$; $N_{\omega} $ is the total number of discrete frequencies considered, therefore leading to a Nyquist (cutoff) frequency of $N_{\omega}\Delta\omega/2$; $\phi_{jl}$ is the independent uniformly distributed random variable in $[0,2\pi]$ generating the stochasticity in the wind dynamic loads; $\Lambda_l$ and $\boldsymbol\Psi_l$ are the $l$th eigenvalue and eigenvector of $\textbf{S}_{\textbf{f}}$; and $\theta_{il} = \tan^{-1}($Im$(\Psi_{il})/$Re$(\Psi_{il}))$ is the complex angle of $\Psi_{il}$. %This explains the process of calibration of the stochastic wind load model to building-specific wind tunnel data.
  
% ------------------------------------------------------------
% ------------------------------------------------------------   

\subsection{Nonlinear Modeling of Steel Structures}
\label{sect:NLM}
\subsubsection{Fiber-based modeling}
\label{sect:FBM}

A fiber-based modeling approach using force-based beam-column elements is adopted in this study based on the recommendations provided in the literature \citep{karamanci2014computational,Uriz2008model}. Large displacement effects are considered by using a corotational formulation \citep{Corot} which enforces an exact nonlinear transformation matrix that captures both the current orientation and length of the element \citep{Corotdescription}. Fiber-based section discretization facilitates the simulation of distributed plasticity and explicit consideration of axial-moment interaction, which is important if large fluctuations in the axial loads are expected during extreme winds. The Menegotto-Pinto material model is used to simulate the hysteretic behavior of steel with combined kinematic and isotropic hardening. The parameters that define the fiber-level stress-strain curve, under monotonic and cyclic loading, for braces are taken from the recommendations provided in \citet{karamanci2014computational}.

% ----------

\subsubsection{Low-cycle fatigue and component buckling}
\label{sect:LCF}

The fiber level strains provide critical information on section plastification and inelastic buckling. They are also key for the prediction of potential component fracture as a result of LCF. In this approach, through the prediction of fiber-level rupture, the potential for section-level fracture is also captured \citep{karamanci2014computational,Urizthesis}. The model initiates fracture based on the Coffin-Manson relationship that expresses the strain amplitude, $\epsilon_i$, at which LCF failure will occur after $N_{fi}$ constant amplitude cycles:
\begin{equation}
\label{eq:LCFeq}
\epsilon_i = \epsilon_0 (N_{fi})^m
\end{equation}
where $\epsilon_0$ is a material parameter that indicates the strain amplitude at which a single complete cycle will cause a fracture; $m$ is a material sensitivity parameter which represents the slope of the Coffin-Manson curve in the log-log space. In general, the wind response of a component will not be at a constant strain amplitude. To account for this, the estimation of the current number of cycles, $n_i(t)$, at a given amplitude, $\epsilon_i$, can be estimated through the modified rain flow cycle counting algorithm outlined in \citet{Urizthesis}. Current damage at the amplitude $\epsilon_i$ can subsequently be estimated by normalizing $n_i(t)$ by the total number of constant amplitude cycles that would be necessary to cause failure, i.e., $N_{fi}$ estimated from Eq. (\ref{eq:LCFeq}). Miner's linear damage accumulation rule can then be used to estimate overall damage as the sum of damage over all amplitude levels: 
\begin{equation}
\label{eq:LCFeq_1}
DI(t) = \sum \frac{n_i(t)}{N_{fi}}
\end{equation}
where $DI(t)$ is the current value of the damage index with $0$ indicating no damage and $1$ indicating failure (i.e. fiber fracture). Importantly for modeling fiber fracture under alongwind type loading, damage due to a monotonic strain increase of $\Delta\epsilon$ can be captured as a half cycle ($n_i = 1/2$) of amplitude $\epsilon_i = \Delta\epsilon$.    
            
Global buckling of a component can be captured through the corotational formulation and considering geometric imperfection by modeling each component with multiple force-based beam column elements and an initial camber \citep{karamanci2014computational,Uriz2008model}. Although local buckling affects the estimation of local inelastic strains, and subsequently, the prediction of section-level fracture, its effect is not considered in this study owing to the use of fiber-based models.

\subsubsection{Damping}
\label{sect:damp}

Rayleigh damping is considered for representing the inherent damping in the structure. The demerits of employing initial stiffness-based Rayleigh damping for inelastic structural systems were emphasized in \citet{Charney2008}. When large inelastic deformations occur and result in the reduction in stiffness and natural frequencies, significant artificial damping may be generated in the lower modes with the initial stiffness approximation. This may lead to overestimation of collapse capacity and consequently, collapse-level reliability. To this end, following the recommendations in \citet{Charney2008}, the damping matrix is formulated based on the tangent stiffness with the proportionality coefficients estimated based on elastic stiffness, $\textbf{K}_0$. That is,
\begin{equation}
\label{eq:dampeq}
\textbf{f}_D (\dot{\textbf{u}}(t),\textbf{u}(t)) = (c_0^{(\textbf{K}_0)}\textbf{M} + c_1^{(\textbf{K}_0)}\textbf{K}_t)\dot{\textbf{u}}(t)
\end{equation}
where $c_0$, $c_1$ are the Rayleigh proportionality coefficients indicated with the superscript $\textbf{K}_0$ to reiterate how they are estimated while $\textbf{K}_t$ is the tangent stiffness matrix computed at the end of the previous time step (last-committed state) during nonlinear dynamic analysis.

% ----------------------------------------------
% ----------------------------------------------

\section{Proposed Collapse Simulation Framework}
\label{SimFrame}

% -------------
\subsection{Integration of Models}
\label{sect:modelinteg}

Further to the adoption of the background numerical models for wind load modeling and nonlinear structural response estimation, the quantification and efficient propagation of input uncertainties in estimating the exceedance probabilities against the predefined limit states of interest is essential to the proposed framework. In particular, if Eq. (\ref{eq:probintg}) is to provide reliability estimates that are comparable with the target reliabilities stipulated in codes and standards, e.g. Table 1.3-1 of the ASCE 7-22 \citep{ASCE7}, a comprehensive set of code-compliant uncertainties in the model parameters, gravity loads, and structural properties must be considered in addition to those of Section ``Wind Load Model'' modeling uncertainty in the wind loads. %As will be seen, these uncertainties can be treated systematically with the background models.

It should be mentioned that in this study, it is recognized that system collapse cannot, in general, be defined explicitly in terms of a set of quantifiable response metrics. Indeed, the occurrence of collapse generally requires the evaluation of a combination of indicators, such as non-convergence of nonlinear analysis, drift levels, the deformed shape at the last converged time step, so as to confirm the onset of a collapse mechanism. As will be illustrated in the case study, the nonlinear modeling environment and stochastic simulation scheme of this study enable such a direct evaluation of system collapse. 
% ---------------------------------------------------------
\subsection{Load Uncertainty}
\label{gravityrvs}

Load uncertainty is arguably the primary source of uncertainty effecting the collapse reliability of a structure. To model wind load uncertainty, the intensity of the wind event is characterized through the site-specific maximum mean hourly wind speed, $\bar{v}_H$, and associated wind direction, $\alpha$, through the models outlined in Section ``Hazard curve and wind directionality''. Record-to-record variability (load stochasticity), on the other hand, is captured through the wind tunnel informed stochastic wind load model of Section ``Load stochasticity''. 

TIn addition, appropriate probabilistic dead, including superimposed, and live load models are required to represent the uncertainty in gravity loads, i.e., the vector $\boldsymbol{\xi}_g$ of Section ``Wind Load and Structural Model Requirements''. To ensure a representation of the live loads that is consistent with how reliability has been traditionally estimated for systems subject to extreme winds, arbitrary point-in-time live loads, $L_{apt}$, should be considered for combination with the stochastic wind loads. In general, the modeling of the uncertainty in the dead, $D$, and arbitrary point-in-time live loads can be directly related to their nominal values \citep{Ellingwood1980,Galambos1982,Zhang_14}. In this respect, the random dead and arbitrary point-in-time live loads of this work are related to their nominal values as reported in Table \ref{tab:DLoad}. 

% ============================================================================================
\begin{table*}
\caption{Basic random variables used for modeling uncertainty in the gravity loads.}
\tabcolsep 2.7pt
\footnotesize
%\centering
\begin{tabular*}{\textwidth}{@{\extracolsep\fill}lllll@{}}\hline
  Parameter & Mean & COV & Distribution & Reference \\\hline
$D$ & 1.05$D_n$ $^{*}$ & 0.1 & Normal & \cite{Galambos1982,Zhang_14} \\
$L_{apt}$ & 0.24$L_{rn}$ $^{*}$ & 0.6& Gamma & \cite{Ellingwood1980,Zhang_14} 
\\\hline
\end{tabular*}  
%\end{center} 
\label{tab:DLoad}     
\small
$^{*}$ $D_n$, $L_{rn}$ : Nominal dead (including superimposed) load and reduced live load.                                                                             
\end{table*}
% ============================================================================================

% -------------

\subsection{Model Uncertainty}
\label{windrvs}

\subsubsection{Wind load model uncertainty}
\label{sec:wlm_unc}

It is important to account for the uncertainties associated with the use of wind tunnel data in calibrating the stochastic wind load model. This can be achieved by multiplying each realization of $\textbf{f}(t)$ with a set of appropriate multiplicative random variables. Accordingly, three unit mean random variables, $w_1, w_2$ and $w_3$, are used to account respectively for the sampling errors due to the finite length of the wind tunnel record, uncertainty due to the use of scale models, and observational errors \citep{Diniz2005probabilistic,Bernardini2015}. These are modeled as truncated normal random variables with a lower bound of zero to ensure non-negativity. In general, the coefficient of variation (COV) for $w_2$ is record length dependent, and a value of 0.075 can be used for a record length of 1-hour or more, at full scale. A summary of these model uncertainties along with their governing distributions is presented in Table \ref{tab:loadrvs}. These uncertainties define the vector $\boldsymbol{\xi}_w$, and are to be considered together with the wind load uncertainties, $\{\bar{v}_H, \alpha, \boldsymbol{\phi}\}$.   

%
% ============================================================================================
\begin{table*}
\caption{Wind load model uncertainties.}
%\begin{center}
\tabcolsep 2.7pt
\footnotesize
%\centering
\begin{tabular*}{\textwidth}{@{\extracolsep\fill}lllll@{}}\hline
  Parameter & Mean & COV & Distribution & Reference \\\hline
$w_{1}$   & 1.0 & 0.075 & Truncated Normal       & \cite{Sadek_04} \\
$w_{2}$   & 1.0 &   0.05  & Truncated Normal       &\cite{Diniz2005probabilistic,Bernardini2015} \\
$w_{3}$   & 1.0 &   0.05  & Truncated Normal       &\cite{Diniz2005probabilistic,Bernardini2015}
\\\hline
\end{tabular*}                                                   
%\end{center} 
\label{tab:loadrvs}                                                 
\end{table*}
% ============================================================================================

% -------------

\subsubsection{Structural model uncertainty}
\label{sec:Systemrvs}

Structural model uncertainty %, $\boldsymbol{\xi}_m$ 
involves geometric imperfections, that will affect onset of member buckling, as well as material parameters that will affect aspects such as onset of yielding, hardening behavior, fatigue and member fracture failure.      

With respect to geometric imperfections, in general, this can be sufficiently modeled by the first buckling mode, given by $\sin{(\pi x)}$ where $x$ denotes the normalized coordinate along the length of the component, that is multiplied by a random scale factor, $\delta$, of random sign \citep{Zhang2016,SHAYAN2014}. With respect to material parameter uncertainty, among the Menegotto-Pinto material model parameters, the Young's modulus $E$, strain hardening ratio $b$, yield strength $F_y$, fatigue material parameter $\epsilon_0$, elastic-to-plastic transition parameter $R_0$, are are all modeled as random variables. The cyclic hardening of steel is influenced by four dimensionless isotropic hardening parameters denoted as $a_1$, $a_2$, $a_3$, and $a_4$, among which $a_1$, $a_3$ are modeled as random variables, consistent with the suggestions outlined in \citet{karamanci2014computational}, while the dimensionless transition parameter, $R_0$, is modeled as a truncated normal random variable with lower and upper bounds of 15 and 25, respectively. To model uncertainty in the damping, the Rayleigh proportionality coefficients, $c_0$ and $c_1$, can be estimated by imposing the first two viscous damping ratios are equal to the lognormal random damping ratio $\zeta$. A summary of the structural model uncertainties, along with their governing distributions, is presented in Table \ref{tab:structrvs}.

It should be observed that initial camber displacements should be generated independently and identically for every member. Similarly, the random variables defining the material model should be independently and identically sampled for each unique structural section of the system. These modeling assumptions ensure material-level independence of strength/stiffness among structural components belonging to different section sizes and independent onset of buckling.

% ============================================================================================
\begin{table*}
\caption{Summary of basic random variables for structural properties.}
%\begin{center}
\tabcolsep 2.7pt
\footnotesize
\begin{tabular*}{\textwidth}{@{\extracolsep\fill}lllll@{}}\hline
  Parameter & Mean & COV & Distribution & Reference \\\hline
$E$  & 200 GPa & 0.04 & Lognormal & \cite{Bartlett2003,Zhang_14} \\
$F_y$ & 1.1$F_{yn}^*$ & 0.06 & Lognormal & \cite{Bartlett2003,Zhang_14} \\
$b$  & 0.001 & 0.01 & Lognormal & \cite{karamanci2014computational,Galambos1978LN} \\
$\epsilon_0$    & 0.077 & 0.161 & Lognormal  & \cite{karamanci2014computational} \\
$R_0$    & 20 & 0.166  & Truncated Normal    &\cite{karamanci2014computational} \\
$a_1$    & 0.01 & 2 & Lognormal      &\cite{karamanci2014computational}  \\
$a_3$   & 0.02 & 0.5 & Lognormal      &\cite{karamanci2014computational}  \\
$\delta$ & 0.000556$L^{**}$ & 0.77 & Normal      &\cite{Zhang2016,SHAYAN2014} \\
$\zeta$   & 0.015 &   0.4  & Lognormal &\cite{Bernardini2015,Kwon2015_DAMPING,Davenport1986damping}
\\\hline
\end{tabular*}                                                   
%\end{center} 
\label{tab:structrvs}  
\small
$^{*}$ $F_{yn}$ : Nominal yield strength.\\
$^{**}$ $L$ : Member length.                                               
\end{table*}
% ============================================================================================
  
% -------------------------------------------- 
% ---------------------------------------------------------
\subsection{Stochastic Simulation Scheme}
\label{sect:simscheme}

The estimation of Eq. (\ref{eq:probintg}) cannot be carried out through direct stochastic simulation, e.g. standard Monte Carlo (MC) methods, as the computational effort associated with NLTHA would fast become prohibitive, especially in light of the need to estimate the small failure probabilities necessary for estimating collapse reliability. To overcome this, an efficient stratified sampling scheme is employed in this study \citep{esrel21}. The approach involves partitioning the probability space into mutually exclusive and collectively exhaustive subspaces, called strata. For a given limit state of interest, $\bar{v}_H$ can be considered as the primary demand variable affecting its violation. The probability space can therefore be stratified in terms of $\bar{v}_H$ by dividing the support of $\bar{v}_H$ into $N_w$ mutually exclusive and collectively exhaustive wind speed intervals (WSIs),  $E^{i}_{\bar{v}_H} = [\bar{v}^L_{H_i},\bar{v}^U_{H_i})$ for $N_w = 1,2,...,N_w$ where $\bar{v}^L_{H_i}$ and $\bar{v}^U_{H_i}$ are the lower and upper bound wind speeds defining the $i$th WSI. To ensure the collectively exhaustive nature of the stratification, the lower bound of the first WSI is taken as zero while the upper bound of the last WSI is taken as infinity. By leveraging the theorem of total probability while using standard MC methods within each strata, the estimator of Eq. (\ref{eq:probintg}) for a given limit state, can be written as:
\begin{equation}
\label{eq:TPTeq}
\hat{\lambda}_\mathcal{R} = \nu \hat{P}_{f_\mathcal{R}} = \nu \sum_{i=1}^{N_w} \hat{P}_{{f_\mathcal{R}} |E^{i}_{\bar{v}_H}} P(E^{i}_{\bar{v}_H})  
\end{equation}
where $\hat{P}_{f_\mathcal{R}}$ is the estimator of the probability of failure given the occurrence of a wind event, $P(E^{i}_{\bar{v}_H})$ is the probability of a wind speed sample belonging to $E^{i}_{\bar{v}_H}$ (which can be directly estimated from the distribution function, $F_{\bar{V}_H}$, of Eq. (\ref{eq:hazardeq})), while $\hat{P}_{{f_\mathcal{R}}|E^{i}_{\bar{v}_H}}$ is the MC estimator of the conditional failure probability: 
\begin{equation}
\label{eq:ConProp}
P_{{f_\mathcal{R}}|E^{i}_{\bar{v}_H}} = \int_{\alpha}  G(\mathcal{R}|\alpha, E^{i}_{\bar{v}_H}) \abs{dG(\alpha|E^{i}_{\bar{v}_H})}
\end{equation}
By observing that wind effects are closely related to the square of $\bar{v}_H$, it is opportune to define the bounds of the WSIs such that they represent intervals of equal squared wind speed difference \citep{esrel21,ZhiCheng20}.

It can be shown that the variance of $\hat{P}_{{f_\mathcal{R}}}$ can be written as:
\begin{equation}
\label{eq:VarRrop}
    \Var(\hat{P}_{f_\mathcal{R}}) = \sum_{i=1}^{N_w} \frac{P_{{f_\mathcal{R}}|E^{i}_{\bar{v}_H}}(1 - P_{{f_\mathcal{R}}|E^{i}_{\bar{v}_H}}) P(E^{i}_{\bar{v}_H})^2}{n_i}
\end{equation}
where $n_i$ is the number of samples used to estimate $P_{{f_\mathcal{R}}|E^{i}_{\bar{v}_H}}$ through MC simulation. Eq. (\ref{eq:VarRrop}) illustrates how, given a total number of MC samples, $N = \sum n_i$, and a limit state $\mathcal{R}$ of interest, the variance of $\hat{P}_{f_\mathcal{R}}$ can be minimized by optimally allocating the samples between the $N_w$ strata. This optimal allocation is central to the reliability assessment framework of this work, as it facilitates the propagation of the uncertainties described in the previous section by rendering the framework computationally feasible. The variance provided by Eq. (\ref{eq:VarRrop}) is also crucial in providing confidence in the estimation of the failure probabilities. The only caveat in applying Eq. (\ref{eq:VarRrop}), is the need to know, \textit{a priori}, $P_{{f_\mathcal{R}}|E^{i}_{\bar{v}_H}}$. To this end, a preliminary analysis can be carried out that provides first estimates of  $P_{{f_\mathcal{R}}|E^{i}_{\bar{v}_H}}$. Further information on the theoretical basis for the application of the above-discussed simulation scheme in the simultaneous estimation of small failure probabilities associated with multiple general limit states can be found in \cite{esrel21}.
% ---------------------------------------------------------

\section{Case Study}
\label{Case}

\subsection{Building Description}
\label{sec:building}

The proposed framework is illustrated on a 2D steel braced frame extracted from a 3D archetype building assumed to be located in New York City, NY. The considered building is based on one of several archetype buildings designed by the ASCE 7-22 Task Committee on PBWE for the advancement of PBWE in practice. The structural system of the building reflects the current state of practice and was based on design checks carried out using equivalent static wind loads (derived from the wind tunnel data that will be used to calibrate the stochastic wind load model) corresponding to a 700-yr return period wind speed and a damping ratio of 2\%. The gravity loads considered in the design comprise a nominal self weight floor load of 2.39 kPa, a superimposed dead load of 0.72 kPa, a cladding weight 1.20 kPa acting on the vertical perimeter surface, and a reduced live load of 1.24 kPa. Perfect correlation is assumed among the dead and live loads at every floor level, consistent with available literature \citep{Zhang_14, Zhang2016}. Nominal Young's modulus, shear modulus and yield strength for the steel are taken as 200 GPa, 77 GPa and 345 MPa respectively.

\begin{figure}
\centering
\includegraphics[scale=0.7]{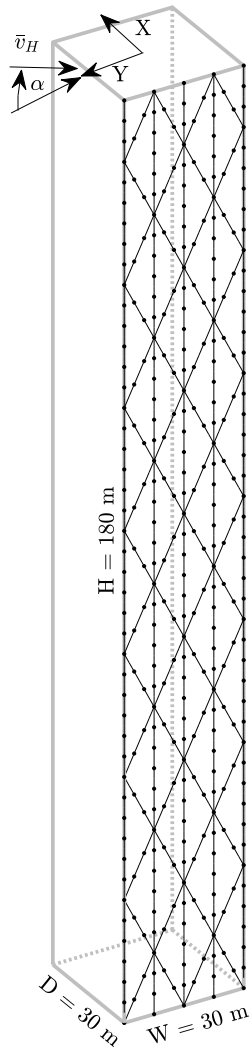}
\caption{The planar frame extracted from the 3D archetype building.}
\label{fig:2Dframe}
\end{figure}

The building is square in plan with a story height is 4 m. The extracted 2D frame is one of the four lateral load resisting perimeter frames, as shown in Fig. \ref{fig:2Dframe}, and is composed of perimeter beams (not shown in the figure), columns, and braces. All beams are designed as pin-ended, elastic, and participate only in carrying gravity load. All columns and braces in the considered frame belong to the American Institute for Steel Construction (AISC) W14 family of wide flange sections. The floor system is composed of 63.5 mm of lightweight concrete placed over a 76.2 mm composite deck.

% ============================================================================================
% ============================================================================================

\subsection{Nonlinear Structural Model}
\label{NLmodel}

The structural modeling and analyses are implemented in OpenSees \citep{OpenSees}. The columns and braces are modeled using force-based beam-column fiber elements, each with five integration points. In discretizing the sections, six fibers are used along the flange width and web depth, while two fibers are used along the flange and web thicknesses \citep{karamanci2014computational}. Section aggregators are used to model shear deformation that is assumed elastic. The LCF model of Section ``Low-cycle fatigue and component buckling'', Menegotto material model and large displacements are implemented through the \textit{Fatigue} model wrapped around \textit{Steel02} and \textit{Corotational} geometric transformation. The floor system is modeled as rigid in its plane. Gravity loads are applied to the columns and braces proportionate to their respective tributary areas, whereas the remaining gravity loads are carried by a leaning column composed of elastic beam-column elements, which is linked to the lateral load resisting frame through moment-released connections. To enable global buckling of a component, as recommended in \citet{karamanci2014computational} and \citet{Uriz2008model}, each component is modeled with two force-based beam-column elements of equal length with initial camber imposed at the connecting node. The limit state of component buckling was assumed to be exceeded if the displacement at the mid-length node was greater than 5\% of the length of the component. Uncertainties in the structural system followed the recommendations outlined in Section ``Structural model uncertainty'' and listed in Table \ref{tab:structrvs}. The dimensionless fatigue model parameter, $m$, and deterministic material parameters, $a_2$ and $a_4$, were taken as constants with values of -0.3, 1, and 1 respectively \citep{karamanci2014computational}. Throughout this section, partially yielded (or fractured) components refer to those with at least one yielded (or fractured) fiber, while fully yielded (or fractured) components refer to those that have undergone section-level yield (or fracture). The first natural frequency of the structure had a mean of 0.2208 Hz and a coefficient of variation (COV) of 0.051, while the second natural frequency had a mean of 0.7426 Hz and a COV of 0.051.

In implementing the nonlinear dynamic analysis, a default time step of 0.06 s was considered. In addition, a dynamic time-stepping scheme was also adopted to resolve potential numerical non-convergence by considering smaller time steps ($dt$ = 0.003 s, and 0.001 s) during the analysis. The occurrences of simulated collapse were confirmed by both numerical non-convergence (for $dt$ = 0.001 s) and the deformed shape at the last converged time step. If the collapse occurrence could not be readily ascertained from the response at the last converged time step, the analysis was repeated while considering a reduced time step triplet of $dt$ = 0.001 s, 0.0005 s, and 0.0001 s, when in the vicinity of the previously non-converged time step. The collapse criterion was therefore the explicit onset of a collapse mechanism. 

% ------------------------

\subsection{Wind Loads}
\label{WLcase}
\subsubsection{Probability distributions for wind speed and direction}
\label{estimatingpd}

\begin{figure}
\centering
%\fbox{
\includegraphics[width=\textwidth]{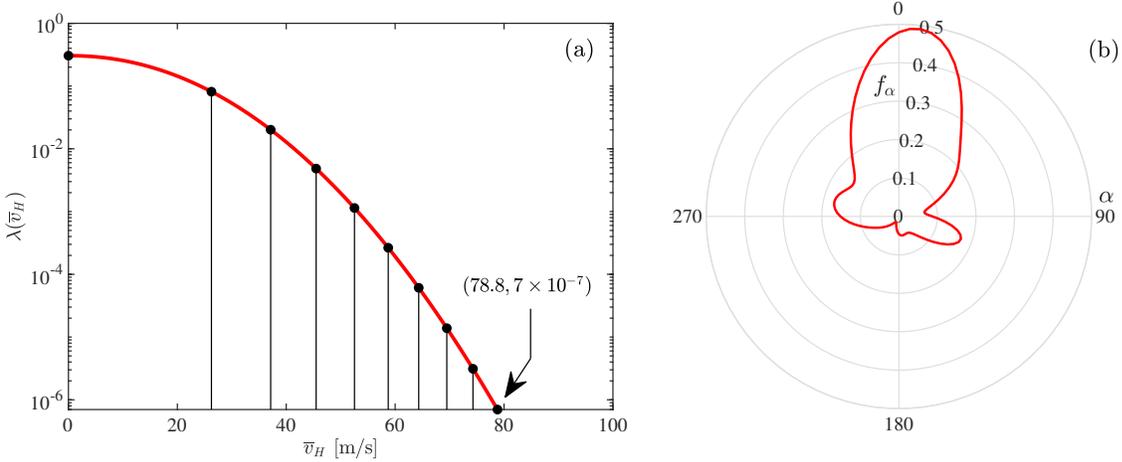}
%}
\caption{(a) Partitioned non-directional hazard curve; (b) Circular PDF of wind direction.}
\label{fig:hazard}
\end{figure}

In order to model the New York City hurricane climate, data from the National Institute of Standards and Technology (NIST) directional wind speed database were utilized. Specifically, the raw wind speed data were transformed to match the building height, $H$ of 180 m, averaging time of 1-hour, and urban terrain. The annual storm frequency, $\nu$, is equal to 0.305 for the considered hurricane data. To model $F_{\bar{V}_H}$ of Eq. (\ref{eq:hazardeq}), a Weibull distribution was considered as it provided point estimates that were consistent with the ASCE 7 wind speeds for New York City. In fitting $F_{\bar{V}_H}$ to the data, maximum likelihood estimates (MLE) of the Weibull parameters were considered. Since unique MLE parameters can be obtained for each realization of the vector, $\textbf{e}$, an average estimate for $F_{\bar{V}_H}$ is considered in this work. The associated wind direction data was used to estimate the marginal distribution of $\alpha$ using a circular kernel density model. In addition to the marginal distributions, the pairwise wind speed and direction samples were used to estimate the copula density function through the adoption of a copula function based on kernel density estimation. Subsequently, the numerical integration of Eq. (\ref{eq:condpdeq2}) was performed to estimate the conditional CCDF of wind direction given wind speed, $G(\alpha|\bar{v}_H)$. Further details about the calculation of the kernel density copula used in the case study can be found in \cite{kdecopula2018}. Fig. \ref{fig:hazard} reports the estimated hazard curve, $\lambda(\bar{v}_H)$, and PDF of wind direction, $f_{\alpha}$.

% ============================================================================================
% ============================================================================================
 
\subsubsection{Calibration of stochastic wind load model}
\label{calibration_pod}

The POD based stochastic model of Section ``Load Stochasticity'' was calibrated to the wind tunnel data provided by Cermak Peterka Petersen (CPP), which were recorded at a sampling frequency of 250 Hz on a 1:400 scale building model for a duration of 81.92 s. A mean reference wind speed of 14.63 m/s was used at a reference height of 0.46 m in the wind tunnel. The time-series data of normalized force coefficients for the two principal horizontal directions (X and Y) and normalized torsional load coefficient at each floor center were recorded for 36 wind directions at $10^{\circ}$ intervals. These were used to calibrate the full-scale wind loads for a total duration of 4200 s, according to Eq. (\ref{eq:PODeq1}), for each simulated wind speed and direction. The first 300 s of each stochastic wind load realization was linearly ramped up to ensure reasonable initial conditions for the NLTHA. The last 300 s was linearly ramped down with the addition of 200 s of zero loading to mark the end of the wind event and provide a means to directly estimate the residual deformations. The first five POD spectral modes were considered in the calibration of the stochastic wind load model. Because of the symmetry of the structural system (i.e., four identical lateral resisting frames with two acting in the X-direction and two acting in the Y-direction), half of the total Y-direction loads on the 3D building were considered to act on the 2D frame. It is assumed that the torsional load acting on the 3D building is resisted through shear forces of equal magnitude acting on each of the four frames. These shear forces, calculated as one-fourth of the torsional load divided by half the building width (i.e., the lever arm), are applied as additional lateral forces on the considered planar frame. It should be observed that this assumption neglects any dynamic effects in torsional loading. However, this is not believed to create a substantial error as the torsional mode of a building of the type considered in this work will in general have a natural frequency that is, not only higher than that of the first two transnational modes, but also towards the higher end of the wind spectrum and therefore only moderately effected by dynamic amplification.

\subsection{Simulation Strategy}
\label{case_sim}

For the implementation of the stochastic simulation scheme, 10 WSIs were considered adequate \citep{ZhiCheng20}, and a total of 1001 samples was considered. The lower bound wind speed of the last WSI was chosen to obtain an annual exceedance rate (AER) of $7\times10^{-7}$. This value represents the target annual failure probability stipulated in ASCE 7-22 \citep{ASCE7} for a risk category II structure corresponding to sudden failure that leads to a widespread progression of damage, and hence relevant to collapse simulation. The partitioning of the wind speed domain is also illustrated in Fig. \ref{fig:hazard}. 
Since standard MC is carried out within each stratum, each simulation point within a stratum considers a realization of all random variables, sampled from their respective probability distributions. This includes wind direction that is sampled conditional on the wind speed following the procedure outlined in Section ``Hazard curve and wind directionality''.

The allocation of MCS samples among the different WSIs was determined through a preliminary assessment using a total of 250 samples. These samples were used to provide estimates of Eq. (\ref{eq:ConProp})  for the limit state of system collapse. The remaining 751 samples were allocated through minimizing Eq. (\ref{eq:VarRrop}) resulting in the sample distribution reported in Table \ref{tab:samplealloc}.

%---------------------------------------------------------------------------------
\begin{table*}
\caption{Sample allocation between the WSIs.}
\tabcolsep 2.7pt                                                          
\footnotesize
\centering
\begin{tabular*}{0.6\textwidth}{@{\extracolsep\fill}llll@{}}\hline
%\begin{tabular*}{llll}\hline
%\hline
WSI & $\bar{v}^L_{H_i}$ [m/s] & $\bar{v}^U_{H_i}$ [m/s] & $n_i$ \\ \hline

1 & 0.00 & 26.26 & 12\\
2 & 26.26 & 37.14 & 12\\
3 & 37.14 & 45.49 & 16\\
4 & 45.49 & 52.53 & 35\\
5 & 52.53 & 58.73 & 553\\
6 & 58.73 & 64.33 & 221\\
7 & 64.33 & 69.49 & 75\\
8 & 69.49 & 74.29 & 35\\
9 & 74.29 & 78.79 & 22\\
10 & 78.79 & $\infty$ & 20\\
\hline
\end{tabular*}
\label{tab:samplealloc}     
\end{table*}
%---------------------------------------------------------------------------------

\subsection{Results}
\label{sec:results}

\subsubsection{Preamble}
\label{preamble}

The results of this study are presented in two parts. The first part focuses on the discussion of observed nonlinear behavior of individual collapse samples with a focus on the type of collapse mechanism, the concurrent violation of component-level limit states, and the influence of wind direction. The second part focuses on the reliability estimates evaluated through Eq. (\ref{eq:Reliability_eq}) and the stratified sampling scheme of Section ``Stochastic Simulation Scheme''.
% ============================================================================================
\begin{figure}
\centering
%\begin{center}
%\fbox{
\includegraphics[scale=1]{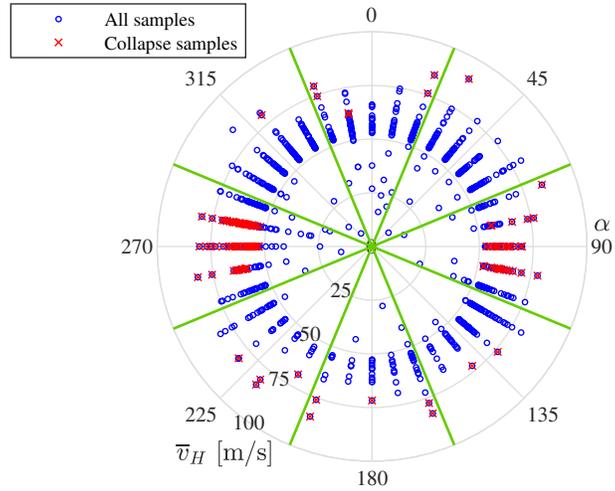}
%}
\caption{Collapse samples as compared to all simulated samples.}
%\end{center} 
\label{fig:collapsedir}
\end{figure}

% ============================================================================================
\subsubsection{Discussion on collapse samples}
\label{sample_results}

A total of 134 collapse samples were identified. Fig. \ref{fig:collapsedir} shows the wind speed and direction pairs, $(\bar{v}_H, \alpha)$, of all simulated wind events while highlighting the pairs in which collapse occurred. From the figure, it is clear that the majority of the collapses occurred in the sectors indicating acrosswind type response (i.e. $\alpha \in [67.5^{\circ}, 112.5^{\circ}] {\displaystyle \cup } [247.5^{\circ}, 292.5^{\circ}]$). In fact, the fraction of collapse samples in these sectors is 86\% in comparison to 8.2\% for sectors associated with an alongwind type response, notwithstanding how the wind speeds are similar in both sectors. This can be attributed to the significantly larger acrosswind aerodynamic responses of the building, which is expected for the square cross-section of the study building. 

% =====================================
\begin{figure}
\centering
%\begin{center}
%\fbox{
\includegraphics[scale=0.6]{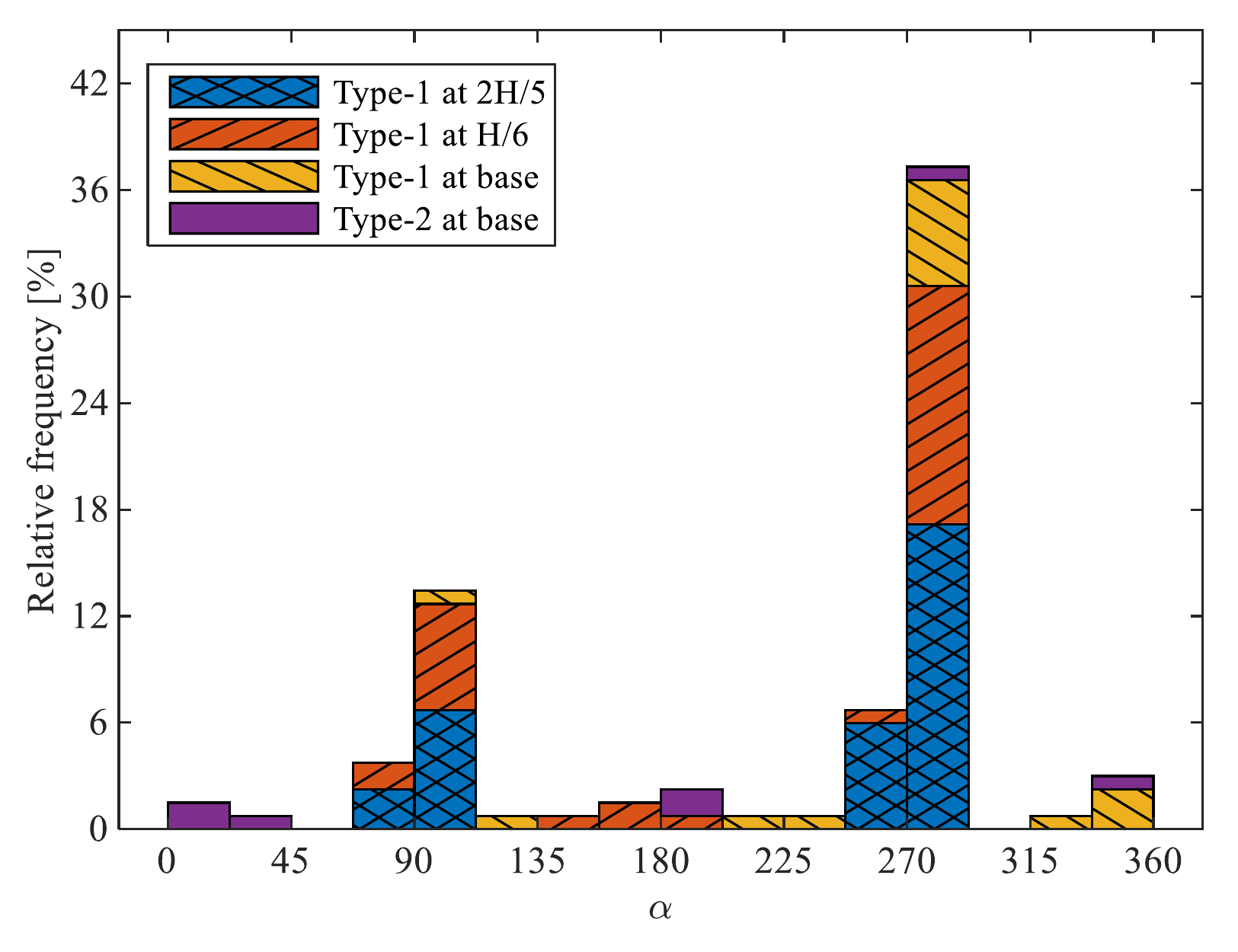}
%}
\caption{Relative frequency of prominent collapse mechanisms and their dependency on wind direction.}
%\end{center} 
\label{fig:collapsefreq}
\end{figure}
% =====================================

The observed collapse mechanisms were classified into 10 types based on the pattern of deformation at the onset of collapse and the location of significant damage. In general, two types of deformation pattern were observed. The first was characterized by an upper block of stories overturning around an area of significant localized damage through a flexure-type mechanism. This will be referred to as a Type-1 mechanism in the following. The other less observed pattern was characterized by excessive inter-story drift at an individual story, indicating a shear-type mechanism that will be referred in the following as a Type-2 mechanism. Collapse characterized by these mechanisms occurred with severe damage at different locations along the height of the structure. Fig. \ref{fig:collapsefreq} reports the relative frequencies of the mechanisms along with the dependence of their occurrences on the wind direction using a stacked bar representation. The results presented show the three most frequently observed Type-1 mechanisms, as well as the most frequently occurring Type-2 mechanism (the other Type-2 mechanisms occurred a negligible number of times). The figure illustrates how alongwind loading only produced Type-1 and Type-2 mechanisms with failure at the base of the structure. The other mechanisms typically resulted from acrosswind loading.
% =========================================
\begin{figure}
\centering
%\fbox{
\includegraphics[scale=0.7]{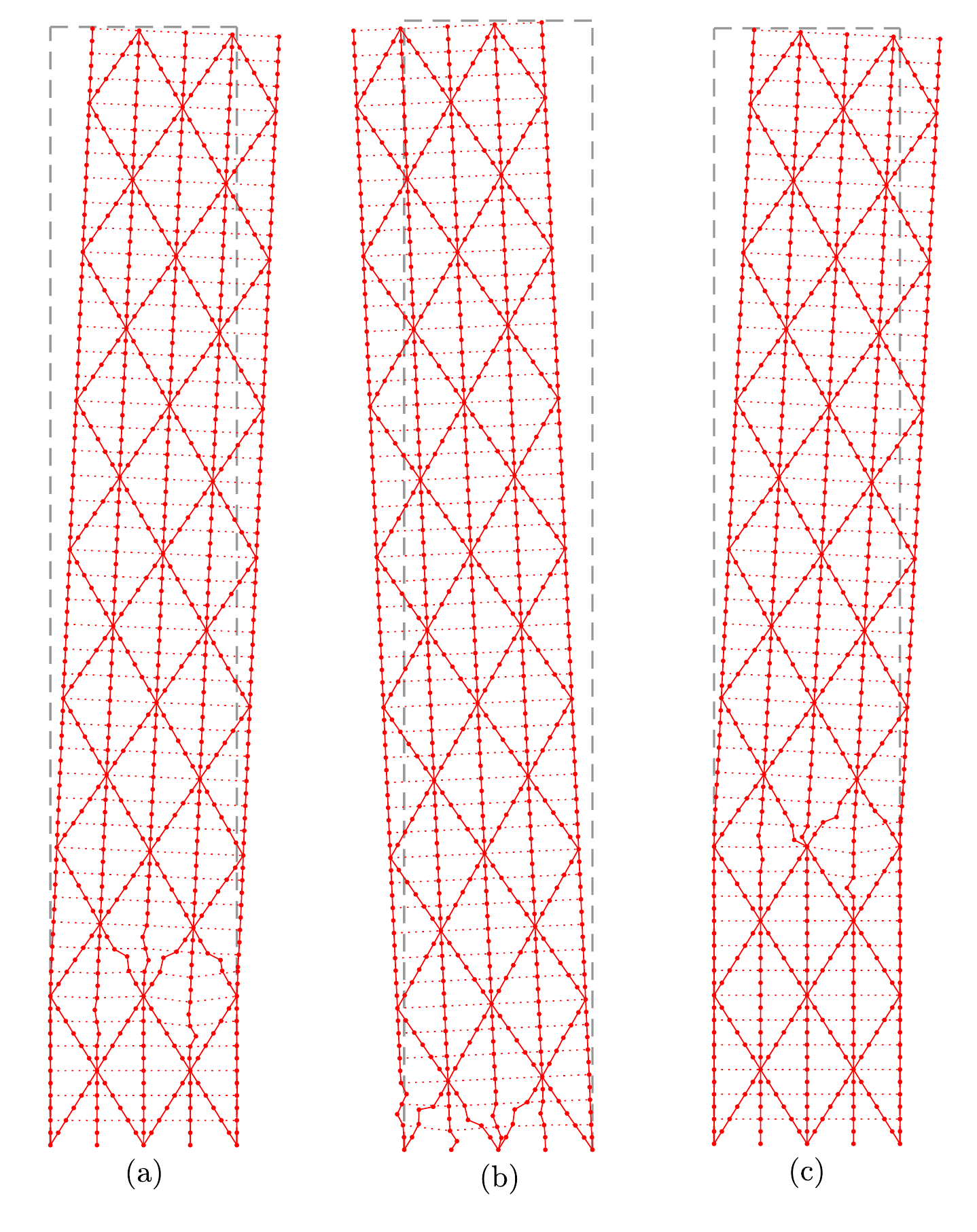}
%}
\caption{Type-1 collapse mechanisms with failure at (a) $H/6$; (b) base; (c) $H/3$.}
\label{fig:flexmech}
\end{figure}
% =========================================

\begin{figure}
\centering
\includegraphics[scale=0.7]{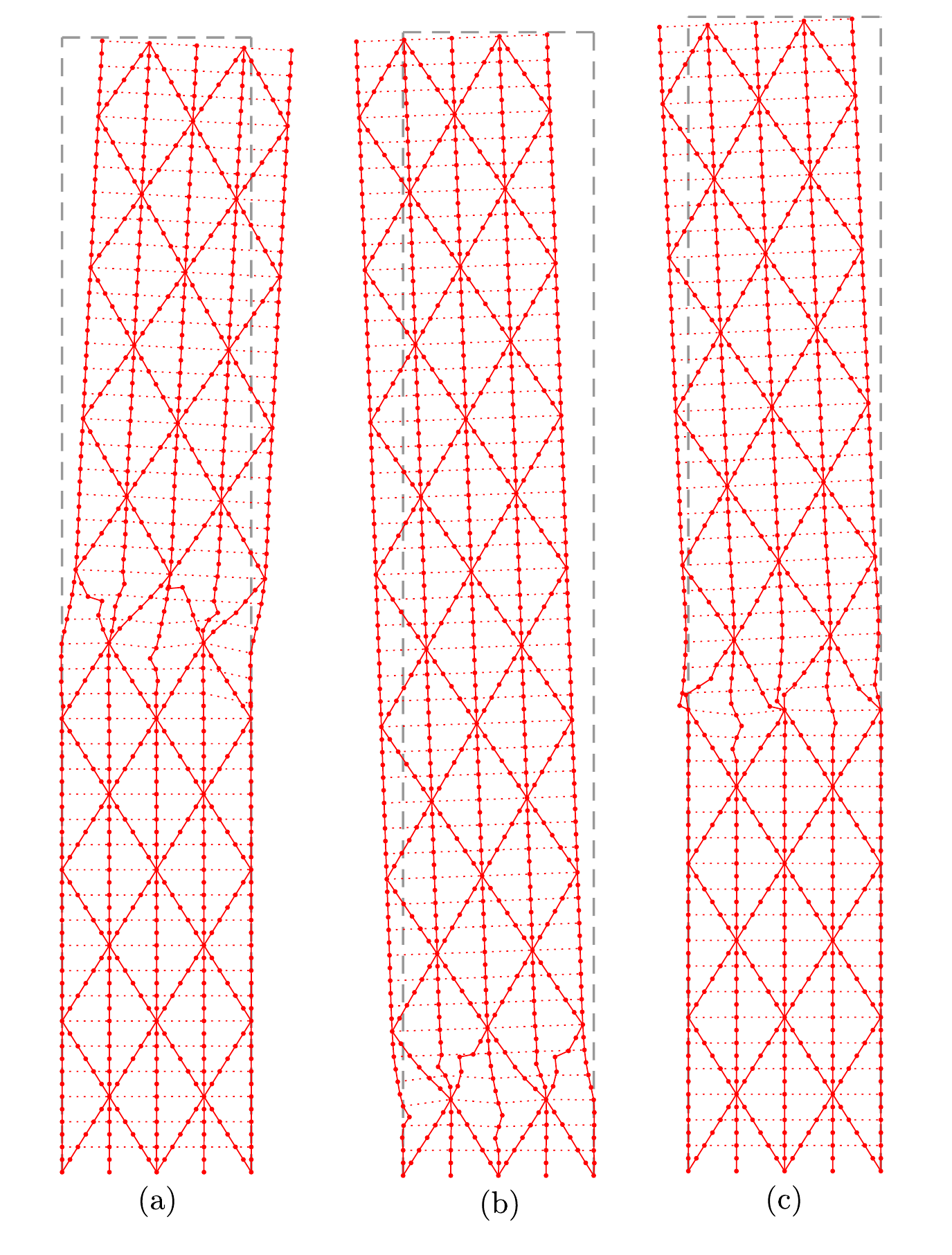}
\caption{Type-2 mechanisms with failure at: (a) $H/2$; (b) base; (c) $2H/5$.}
\label{fig:shearmech}
\end{figure}

\begin{figure}[!htb]
\centering
%\fbox{
	\includegraphics[scale=1]{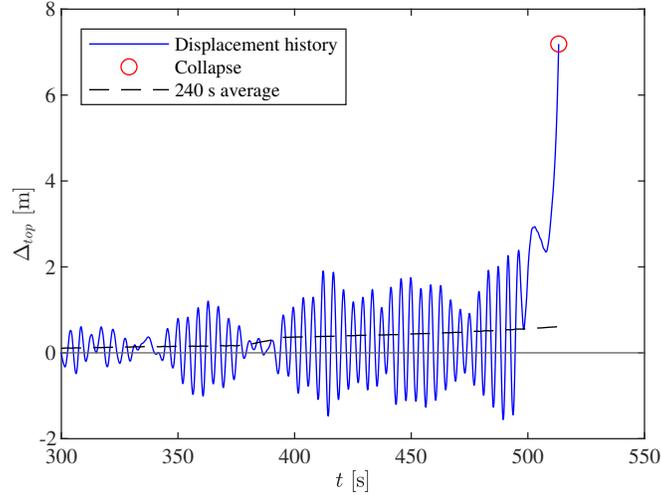}
	%}
	\caption{Collapse sample 1 ($\bar{v}_H = 60.26$ m/s, $\alpha = 280^{\circ}$): Roof displacement time history.}
	\label{fig:cm1disp}
\end{figure}

\begin{figure}[!htb]
\centering
%\fbox{
\includegraphics[scale=0.8]{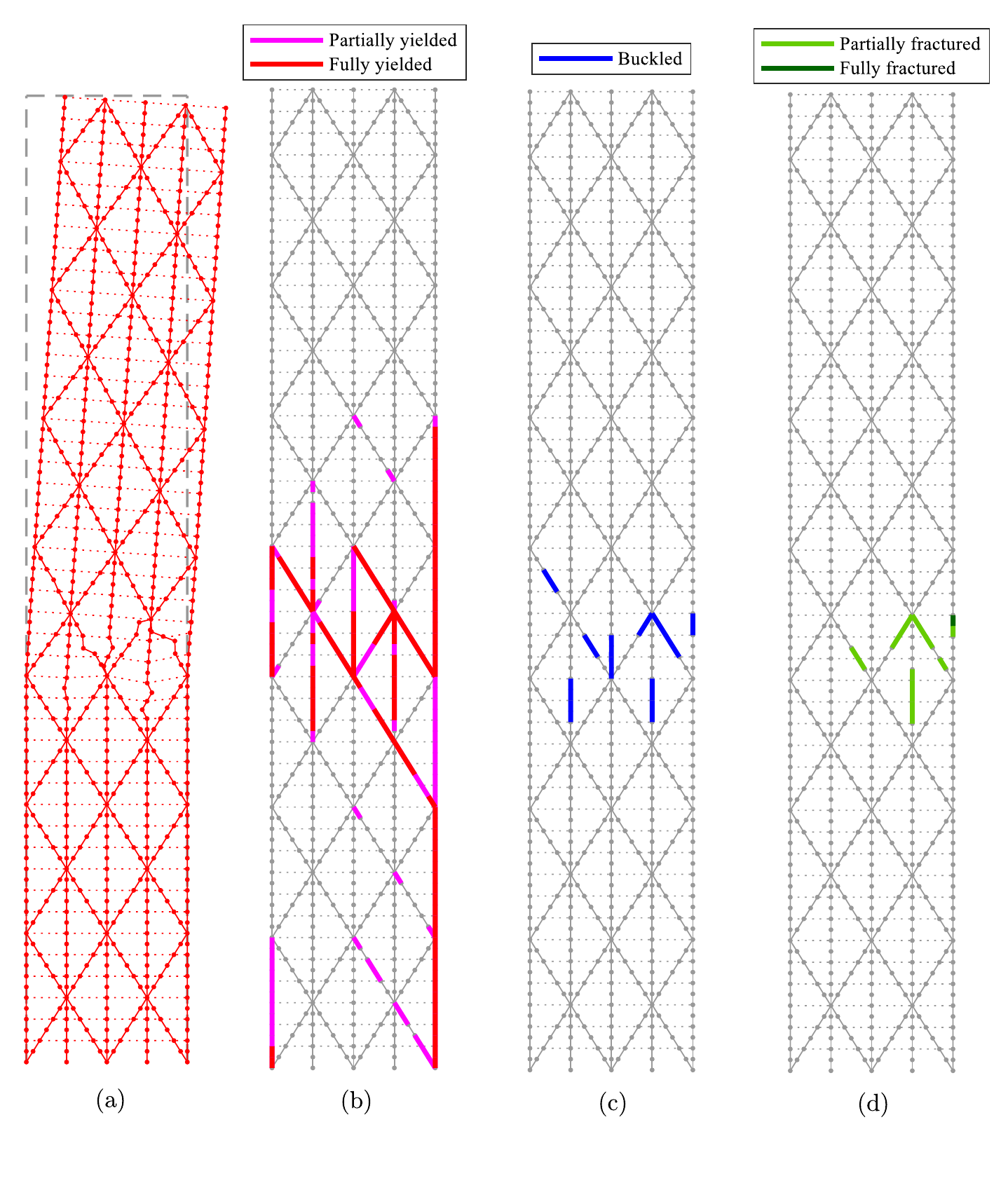}
%}
\caption{Collapse sample 1 ($\bar{v}_H = 60.26$ m/s, $\alpha = 280^{\circ}$): (a) Deformed shape at collapse; (b) Partially and fully yielded components; (c) Buckled components; (d) Partially and fully fractured components.}
    \label{fig:cms1}
\end{figure}

\begin{figure}[!htb]
    \centering
    %\fbox{
    \includegraphics[scale=0.85]{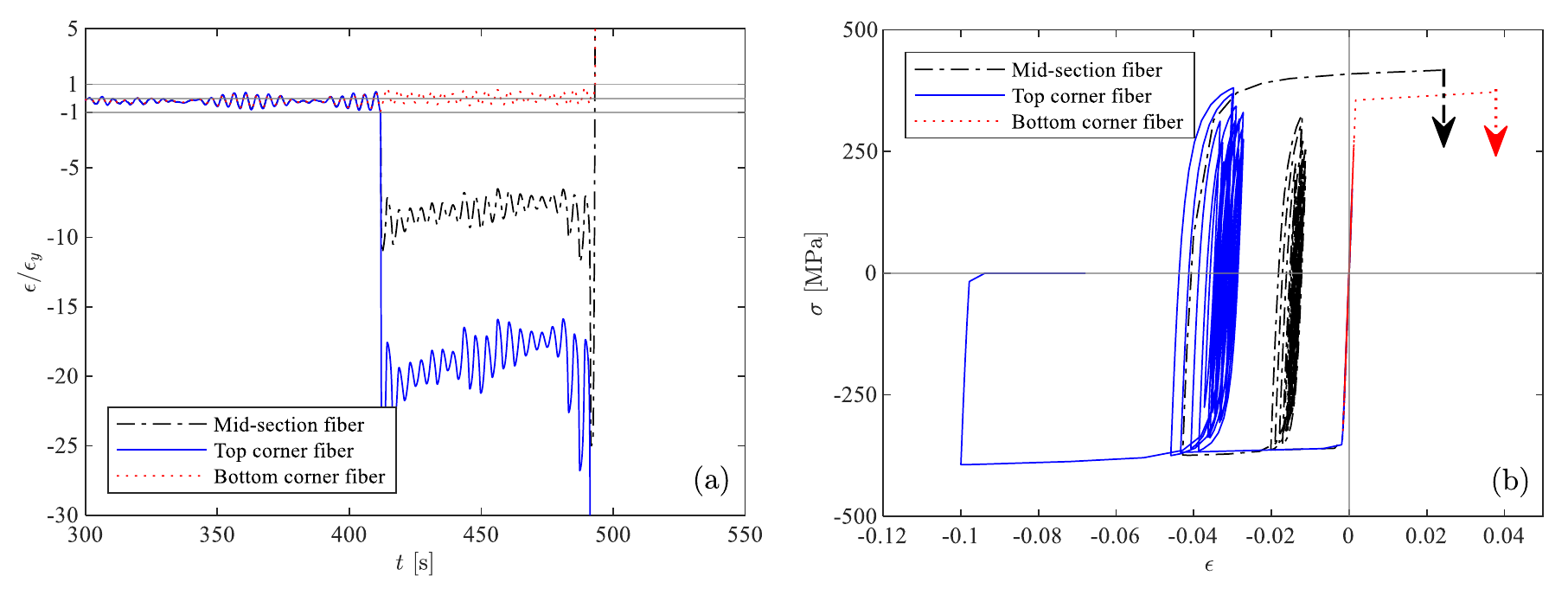}
    %}
    \caption{Collapse sample 1 ($\bar{v}_H = 60.26$ m/s, $\alpha = 280^{\circ}$): (a) Strain histories of fibers belonging to the fully fractured column; (b) Stress-strain histories of the fractured fibers.}
    \label{fig:cm1fibers}
    \end{figure}

Typical instances of Type-1 mechanisms at the onset of collapse are illustrated in Fig. \ref{fig:flexmech} using the deformed shapes (recorded at the last converged step of the analysis) of representative samples. Similarly, Type-2 collapse instances are shown in Fig. \ref{fig:shearmech}. To illustrate the most frequently observed collapse mechanism type, Type 1 with failure at $2H/5$, a representative sample is discussed in detail. In particular, the wind speed and direction were $\bar{v}_H = 60.26$ m/s and $\alpha = 280^{\circ}$, indicating acrosswind loading as reflected by the roof displacement time history of Fig. \ref{fig:cm1disp}. Non-stationarity in the response due to damage accumulation can be observed from the 240 s moving average, where the first 300 s of the response history was removed as it is associated with the load ramp. Fig. \ref{fig:cms1}(a) reports the deformed shape at the onset of collapse while Fig. \ref{fig:cms1}(b) reports the partially and completely yielded components, illustrating how widespread yielding is expected at collapse even outside the area of significant damage accumulation (i.e., at $2H/5$). Figs. \ref{fig:cms1}(c)-(d) report the components that have experienced buckling and/or partial to complete fatigue-induced failure. From Figs. \ref{fig:cms1}(c)-(d), it can be observed that the onset of the mechanism is closely aligned with component failure due to buckling or fatigue. This highlights the importance of capturing such failure modes in the collapse simulation of steel structures subject to extreme winds. The fiber strain histories of two corner fibers and one mid-section fiber of the fully fractured column (as indicated in \ref{fig:cms1}(d)) are reported in Fig. \ref{fig:cm1fibers}(a). The figure indicates the time at which the fibers initially yielded as well as the time of fiber failure (close to the time of system collapse) due to progressive low cycle fatigue-induced damage accumulation. It is interesting to note that the bottom corner fiber, unlike the other two fibers, remained elastic almost until system collapse, closer to which it fractured following a sudden burst of load transfer from the other failed fibers in the section. The stress-strain histories of these fibers are reported in Fig. \ref{fig:cm1fibers}(b), which support the previous observation and illustrate the hysteretic behavior prior to fiber fracture. As illustrated by the figure, when the fatigue life of a fiber is exhausted, the \textit{Fatigue} model drops the stress level to 0 (indicated with downward arrows), however, when the failure is triggered in compression, the stress is dropped at the subsequent zero-force crossing.
% =========================================
% =========================================
% =========================================

% =========================================================

\begin{figure}[!htb]
    \centering
	\includegraphics[scale=0.9]{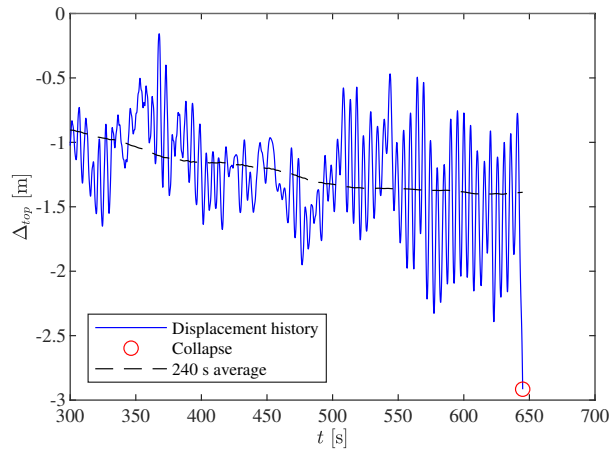}
	\caption{Collapse sample 2 ($\bar{v}_H = 84.93$ m/s, $\alpha = 20^{\circ}$): Roof displacement history.}
	\label{fig:cm2disp}
\end{figure}

\begin{figure}[!htb]
\centering
%\fbox{
\includegraphics[scale=0.85]{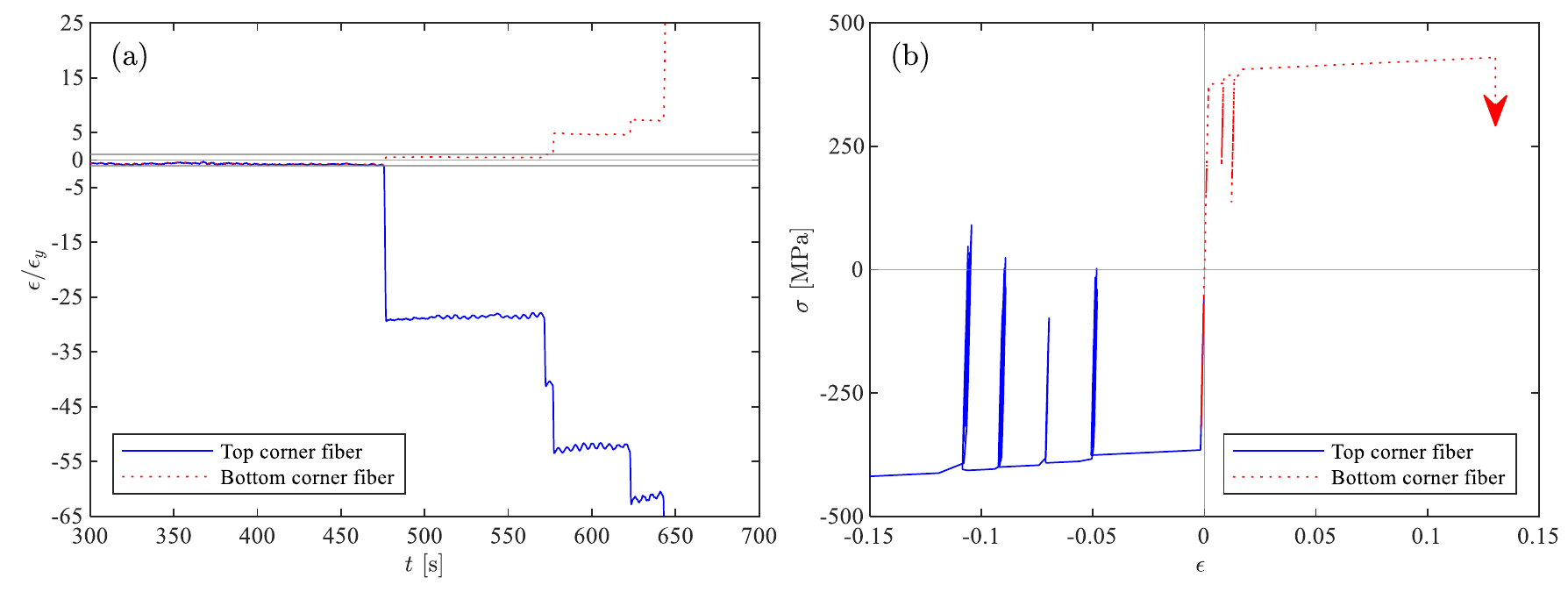}
%}
    \caption{Collapse sample 2 ($\bar{v}_H = 84.93$ m/s, $\alpha = 20^{\circ}$): (a) Strain histories of fibers belonging to a partially brace; (b) Stress-strain histories of the fractured fibers.}
    \label{fig:cm2fibers}
    \end{figure}

To illustrate typical alongwind collapse behavior, a collapse sample with $\bar{v}_H = 84.93$ m/s and $\alpha = 20^{\circ}$ is discussed. Fig. \ref{fig:cm2disp} reports the roof displacement time history for the sample where the characteristic non-zero mean response, as a result of the mean loading component, is clearly visible. Further, accumulation of damage, similar to that seen in Fig. \ref{fig:cm1disp}, is apparent from the change in the 240 s moving average. The presence of a mean loading component also affects the response of the fibers that are expected to yield in a single direction although with stress fluctuations. This phenomenon of unidirectional plastic strain accumulation is characteristic of alongwind response and is known as ratcheting. The strain time histories and stress-strain histories of illustrative fractured fibers are shown in Fig. \ref{fig:cm2fibers}. The figures corroborate the previous statement by showing that fully-reversed hysteretic cycles are absent, and fatigue-induced fracture typically occurs only due to a series of large half-cycle strain fluctuations. 

Out of the ten types of observed mechanisms, four Type-1 mechanisms (failures at the base, $H/6$, $H/3$ and $2H/5$) are illustrated in Fig. \ref{fig:flexmech}, \ref{fig:cms1}a, and three Type-2 mechanisms (failures at the base, $2H/5$ and $H/2$) are illustrated in Fig. \ref{fig:shearmech}. These seven types illustrated in the figures represent 90\% of all the observed mechanisms. The remaining three mechanisms comprise Type-2 failure at $H/3$, and Type-1 failure at $H/2$ and $2H/3$.
% =========================================================
% =========================================

% ===========================================
% =========================================================
\begin{figure}
	\centering
	%\fbox{
	\includegraphics[scale=0.8]{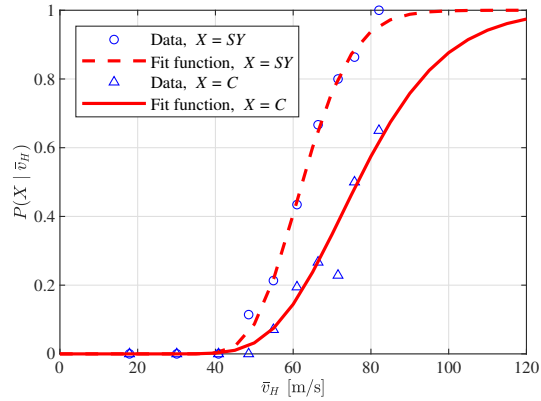}
	\caption{System first yield and collapse fragility curves.}
	\label{fig:cyfragcurves}
\end{figure}
\begin{figure}
	\centering
	%\fbox{
	\includegraphics[scale=0.8]{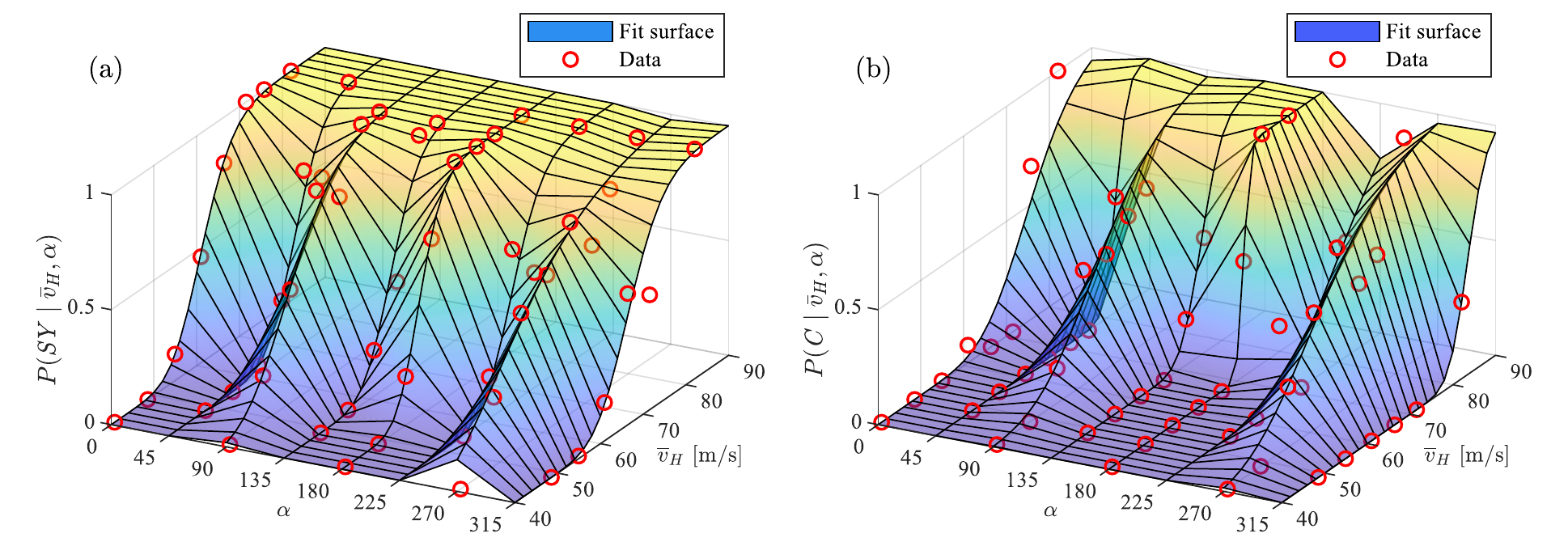}
	\caption{Fragility surfaces for: (a) system first yield; (b) collapse.}
	\label{fig:cyfragsurf}
\end{figure}
% =====================================================

% ============================================================================================

\subsubsection{Reliability assessment}
\label{sec: reliabilityresults}

By implementing the simulation strategy of Section ``Stochastic Simulation Scheme'', the annual failure probabilities for each limit state of interest were estimated. As reported in Table \ref{tab:summary_prob}, the failure probabilities are expressed as AERs and 50-year reliability indices, $\beta_{50}$. From the component limit state reliabilities, it can be observed that the system is somewhat underdesigned and roughly meets the ASCE 7-22 target reliabilities (Table 1.3-1) for a risk category I building. This is not seen as limiting to the discussions of this study that are focused on the relative difference between the reliabilities of the various limit states. As expected, the reliability of system-level first yield is less than that of component-level first yield since the yielding of any one component counts as system yield. It is interesting to observe that the component-level reliabilities against buckling and fracture are around a unit reliability index larger than first yield. This difference is consistent with Table 1.3-1 of the ASCE 7-22 \citep{ASCE7} for limit states that are not expected to be sudden or lead to widespread progression of damage (first yield) vs limit states that are expected to be sudden and lead to widespread progression of damage (component buckling and/or fracture). Of particular interest is the relative comparison between the reliability against system collapse and the reliabilities against component-level yielding, buckling, and fracture. As can be seen, collapse reliability is notably larger than that of first yield (around 0.6) but significantly smaller than that of component failure (around 0.5). These differences suggest that: 1) significant capacity, in terms of reliability, exists post yield; and 2) system collapse can be initiated by the failure of a diverse set of components (i.e., system failure does not require the failure of the same set of components). The first point reinforces the recent interest in leveraging inelasticity in design against extreme winds (as documented by the ASCE Prestandard on PBWD \citep{Prestandard19}), while the second point indicates how component failure is not a good proxy for system failure.
% =========================================
\begin{table*}
\caption{Annual failure rates and 50 year reliability indices.}
\tabcolsep 2.7pt   
\centering
\begin{tabular*}{0.8\textwidth}{@{\extracolsep\fill}llll@{}}
\hline
Limit State & $\hat{\lambda}_\mathcal{R}$ & COV$(\hat{\lambda}_\mathcal{R})$ & $\beta_{50}$ \\\hline
System collapse & $1.18 \times 10^{-4}$ & 9.5\% & 2.52 \\
Component first yield & $5.64 \times 10^{-4}$ & 31.1\% & 1.91 \\
System first yield & $7.40 \times 10^{-4}$ & 26.9\% & 1.79 \\
Component buckling & $3.87 \times 10^{-5}$ & 17.9\% & 2.89 \\
Component fracture & $1.75 \times 10^{-5}$ & 26.5\% & 3.13 \\
\hline
\end{tabular*}                                                   
\label{tab:summary_prob}                                                 
\end{table*}
% =========================================

It is often useful to express performance in terms of fragility functions that directly relate the probability of exceeding a limit state of interest to a measure of hazard intensity, e.g., $\bar{v}_H$. The stratified sampling scheme of Section ``Stochastic Simulation Scheme'' is useful to this end as it provides direct estimates of the failure probabilities for all limit states of interest conditional on each WSI (i.e., intensity measure interval). Lognormal fragility curves can then be defined by first assuming each point estimate of the conditional failure probabilities to be located at the center of the associated WSI, and secondly, applying the maximum likelihood approach for fitting \citep{baker2015}. Following this approach, Fig. \ref{fig:cyfragcurves} shows the fragility curves for collapse ($X = C$) as well as system-level first yield ($X = SY$). The median of the collapse fragility curve is 76.6 m/s, while the dispersion is 0.23. Interestingly, this dispersion is relatively consistent with that observed from collapse analysis of structures subject to seismic actions \citep{FEMA_P_58} before additional dispersion is added to account for sources of uncertainty that are not explicitly modeled, e.g., ambiguity in building definition and construction quality as well uncertainty in the quality and completeness of the analysis model. For system-level first yield, the median and dispersion values of the lognormal fragility curve are 62.4 m/s and 0.16. As would be expected, there is a smaller uncertainty in the onset of yielding as compared to the onset of collapse. In addition to the fragility curves, it is interesting to consider the associated fragility surfaces as they provide a means to directly observe the effects of wind direction. In this respect, Fig. \ref{fig:cyfragsurf} shows the fragility surfaces, constructed using a lognormal assumption for modeling wind speed dependency in each $45^{\circ}$ wind sector, for system first yield and collapse. From the fragility surfaces, the sensitivity of the structure to system first yield and collapse for acrosswind type loading can be clearly seen. 

\begin{figure}
\centering
\includegraphics[scale=0.65]{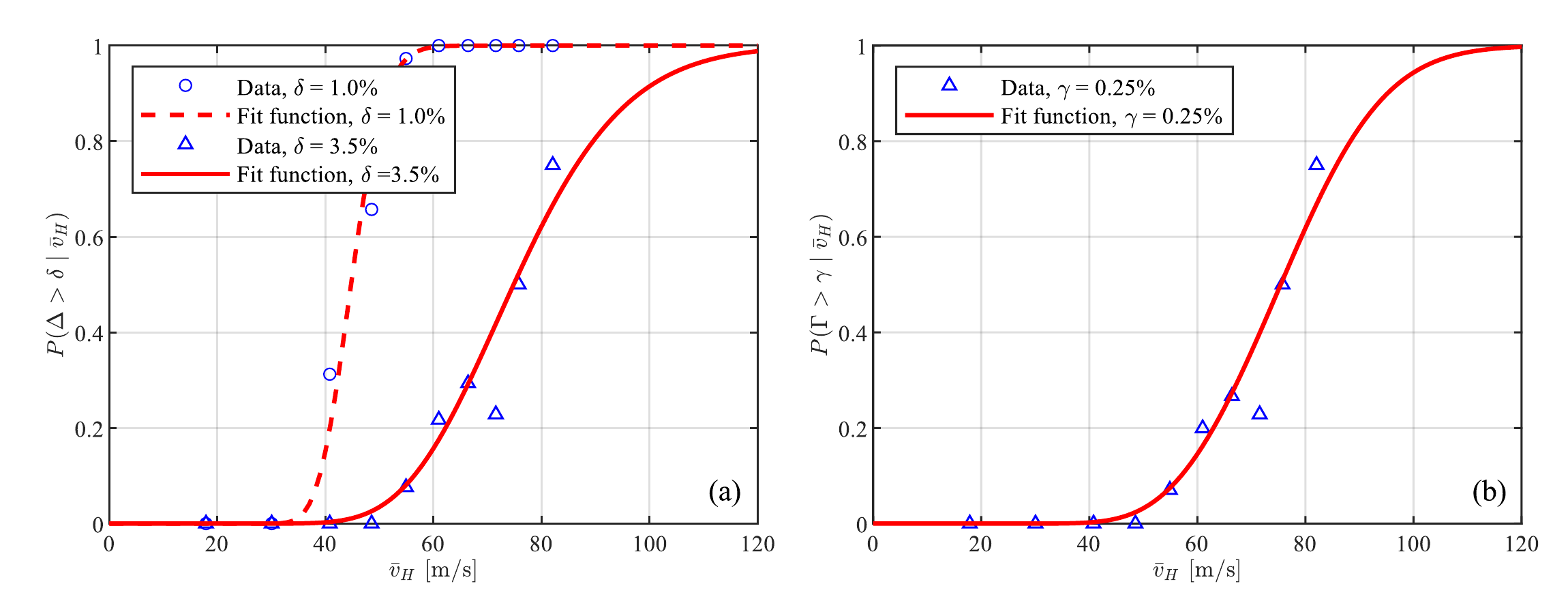}
\caption{Fragility curves for (a) peak story drift ratio; (b) residual story drift ratio.}
\label{fig:idrfragcurve}
\end{figure}

\begin{figure}
\centering
\includegraphics[scale=0.8]{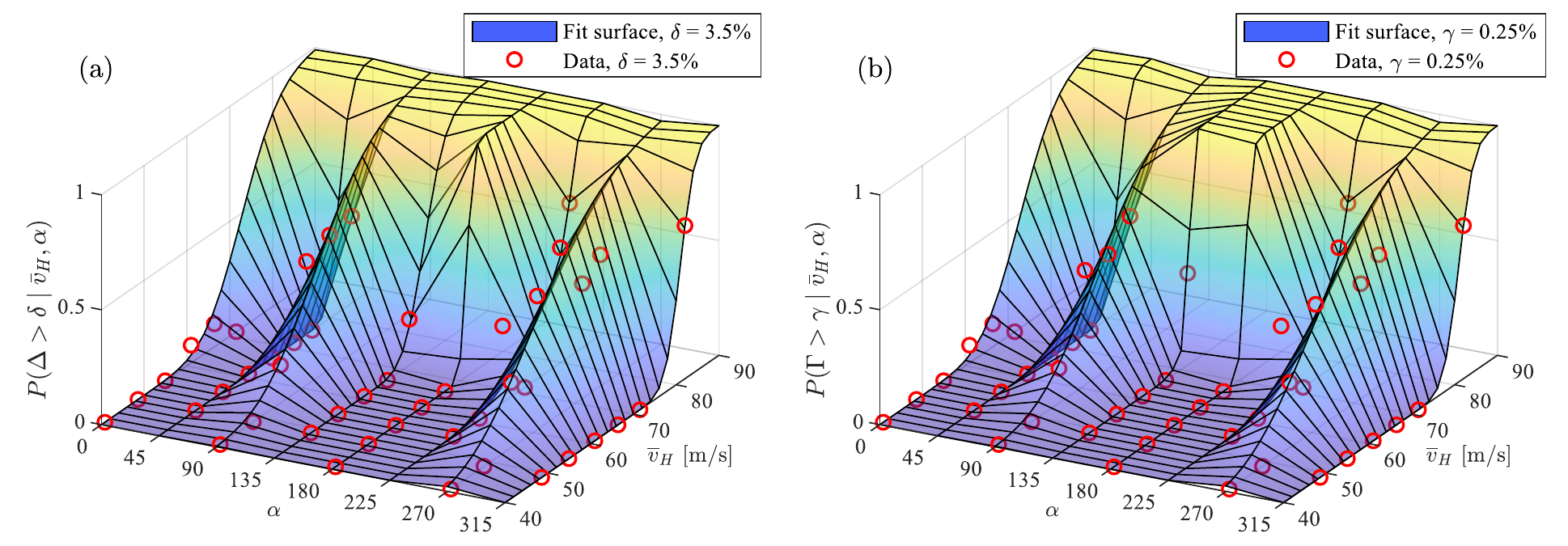}
\caption{Fragility surfaces: for (a) peak story drift ratio; (b) residual story drift ratio.}
\label{fig:idrfragsurf}
\end{figure}

In the context of inelastic design against extreme winds, the degree of sustained structural damage is an important consideration that can be inferred through residual and peak story drift ratios. To this end, fragility curves/surfaces can be developed using the procedure outlined above. The resulting fragility curves for peak story drift $\Delta$, exceeding a threshold $\delta$, and residual story drift $\Gamma$, exceeding a threshold $\gamma$, are reported in Fig. \ref{fig:idrfragcurve}. Notably, the fragility curve of Fig. \ref{fig:idrfragcurve}(a) for $\Delta > 3.5\%$, has a similar shape (roughly the same lognormal distribution parameters) as the system collapse fragility curve, indicating how a 3.5\% peak story drift threshold is a proxy to first collapse. It is interesting to compare the 3.5\% peak story drift fragility curve to that for  1.0 \% (typical design target for peak story drift at continuous occupancy \citep{Prestandard19}). The significant distance between the fragility curves is an indication of the reserves in terms of drift before collapse is expected to occur. With respect to residual drift, it is interesting to observe from Fig. \ref{fig:idrfragcurve}(b) that for a threshold of $\gamma = 0.25\%$, the fragility curve is again similar to the collapse fragility curve. This threshold meets target of 0.5\% suggested in \citet{Prestandard19} for incipient collapse revealing how residual deformation is unlikely to be a concern for this structural system, i.e., the structure is more susceptible to collapse during extreme winds rather than large residual deformation. The fragility surfaces corresponding to the peak and residual story drifts are shown in Fig. \ref{fig:idrfragsurf}. As seen for system first yield and collapse, the sensitivity of the structure to acrosswind type loading is again evident.

% ===========================================
% ===========================================
% ===========================================
% ===========================================

Before closing this section, noteworthy are the remarkably small COV, reported in Table \ref{tab:summary_prob}, achieved for the limit state of system collapse through the stochastic simulation scheme of Section ``Stochastic Simulation Scheme'' with only 1001 samples. To further appreciate the efficiency of the scheme, it should be observed that the number of standard MC samples that would be required to estimate the collapse probability with the same COV, i.e., $9.5\%$, would be around 284,000. This not only illustrates the potential of the scheme, but also underlines the necessity of variance reduction methods in enabling reliability-based collapse assessment of structures subject to extreme winds if excessive computational resources are to be avoided. Finally, it is important to note that the problems of interest to this work involve very high-dimensional spaces of uncertain parameters (order of thousands due to both model and load uncertainty, including the uncertainties necessary to model the stochasticity in the dynamic wind loads), as well as the simultaneous estimation of failure probabilities associated with multiple limit states. These characteristics make the use of methods such as importance sampling extremely challenging as these approaches are not generally applicable to high-dimensional uncertain spaces or the simultaneous treatment of multiple limit states.

\newpage

\section{Summary and Conclusion}
\label{Conclusion}

In this paper, a probabilistic framework is proposed for the characterization of the nonlinear behavior of wind-excited steel structures, with particular focus on the estimation of the annual rate of inelastic excursions (system-level yielding) and reliability against component and system-level collapse. The framework is based on integrating a high-fidelity fiber-based structural modeling environment, capable of capturing nonlinear phenomenon associated with yielding, buckling, low-cycle fatigue and component fracture, with a stratified stochastic simulation scheme for propagating general uncertainty, including stochastic variability in the dynamic winds loads, in estimating rare event probabilities using small sample sets.    

The framework was illustrated on an archetype steel structure that is representative of current design practice from which key insights into the collapse behavior of steel structures subject to extreme winds were gained. In particular, the distinct difference between the acrosswind and alongwind behavior was observed in regard to collapse susceptibility as was the significant variability in the collapse mechanisms and the importance of buckling and low-cycle fatigue for initiating collapse. The stress-strain histories also revealed the difference in the nature of low-cycle fatigue for acrosswind and alongwind loading with acrosswind loading producing behavior similar to that seen for seismic loading and alongwind loading producing responses characterized by significant ratcheting. System-level collapse reliability was seen to be notably lower than component-level failure suggesting the need for either explicit collapse simulation or the use of carefully developed wind collapse fragility functions when evaluating the performance of wind-excited structural systems designed to experience controlled inelasticity.       
% ============================================================================================

\section*{Acknowledgments}
This research effort was supported in part by the National Science Foundation (NSF) under Grant No. CMMI-1750339 and CMMI-2118488. This support is gratefully acknowledged.

\section{Data Availability Statement}
All data, models, and code generated or used during the study appear in the submitted article.

%\section{References}
{\small
\bibliography{Bibliography}
}
\end{document}